\documentclass[useAMS,usenatbib]{mn2e}

\usepackage{amsmath}
\usepackage{amssymb}
\usepackage{graphicx}
\usepackage{aas_macros}

\title[Importance of initial conditions for SF]{Importance of the initial conditions for star formation –-- II. Fragmentation-induced starvation and accretion shielding}
\author[Girichidis et al.]{Philipp~Girichidis$^{1,2,3}$\thanks{email: \texttt{philipp@girichidis.com}}, Christoph~Federrath$^{1,4,5}$, Robi~Banerjee$^{3}$, \& Ralf~S.~Klessen$^1$
\vspace*{0.2cm} \\
\scriptsize
$^1$Zentrum f\"ur Astronomie der Universit\"at Heidelberg, Institut f\"ur Theoretische Astrophysik, Albert-Ueberle-Str.~2, 69120 Heidelberg, Germany \\
\scriptsize
$^2$Cardiff School of Physics and Astronomy, The Parade, Cardiff, CF24 3AA, UK \\
\scriptsize
$^3$Hamburger Sternwarte, Gojenbergsweg 112, 21029 Hamburg, Germany\\
\scriptsize
$^4$Ecole Normale Sup\'{e}rieure de Lyon, CRAL, 69364 Lyon Cedex 07, France\\
\scriptsize
$^5$Monash Centre for Astrophysics (MoCA), School of Mathematical Sciences, Monash University, Vic 3800, Australia
}

\newcommand{\ColorBar}{
  \vspace{0.2cm}
  \includegraphics[width=8cm]{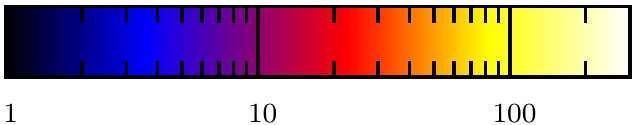}\\
  column density [g~cm$^{-2}$]\\\vspace{0.2cm}
}

\newcommand{\rhoav}{\langle\rho\rangle}
\newcommand{\rkl}[1]{\left(#1\right)}

\newcommand{\skl}[1]{\left\langle#1\right\rangle}
\newcommand{\phn}{\phantom{0}}
\newcommand{\phnn}{\phantom{0}\phantom{0}}

\begin{document}

\maketitle
\begin{abstract}
We investigate the impact of different initial conditions for the initial density profile and the initial turbulence on the formation process of protostellar clusters. We study the collapse of dense molecular cloud cores with three-dimensional adaptive mesh refinement simulations. We focus our discussion on the distribution of the gas among the protostellar objects in the turbulent dynamical cores. Despite the large variations in the initial configurations and the resulting gas and cluster morphology we find that all stellar clusters follow a very similar gas accretion behaviour. Once secondary protostars begin to form, the central region of a cluster is efficiently shielded from further accretion. Hence, objects located close to the centre are starved of material, as indicated by a strong decrease of the central accretion rate. This Fragmentation Induced Starvation occurs not only in rotationally supported discs and filaments, but also in more spherically symmetric clusters with complex chaotic motions.
\end{abstract}


\section{Introduction}
The current paradigm of star formation suggests that most stars form in groups rather than in isolation \citep{Lada03}. However, the fraction of stars that form in a bound clustered environment is still a matter of debate and varies between different models \citep{BressertEtAl2010}. Whereas massive stars only seem to form in a dense environment, low-mass stars form in all observed star-forming regions. Numerical simulations suggest the formation of massive stars simultaneously with the formation of a cluster \citep{Smith09,Peters10a,Peters10b,Peters10c}. The spatial distribution of stars in the forming cluster shows a dependence on the stellar mass: more massive stars are located closer to the centre of the group, low-mass stars tend to populate the outer regions. Whether this mass segregation is primordial and therefore determined by the formation scenario or due to a dynamical relaxation process is still a matter of debate and might differ significantly among different clusters \citep[see, e.g., the reviews by][]{MacLow04,McKee07,Zinnecker07}. Commonly found in observations as well as in numerical simulations is a universal distribution of masses (initial mass function, IMF, \citealt{Scalo86}, \citealt{Scalo98}, \citealt{Kroupa01}, \citealt{Chabrier03}, \citealt{BastianEtAl2010}). Yet it is unclear, what is the influence of various physical processes and initial conditions on the initial mass distribution in star forming regions. Different physical processes like radiative feedback and magnetic fields are likely to have an impact on the collapse of gas clouds  and the subsequent formation of a stellar cluster. Radiative feedback from protostars can have two opposite effects. On the one hand, it heats the surrounding gas and thus contributes to stabilising the gas against collapse by increasing the Jeans mass. In case of massive protostars, the strong contribution in UV ionises the gas, forming \textsc{Hii} regions. Hydrodynamic simulations of forming clusters show that radiative feedback tends to reduce the degree of fragmentation, but does not suppress it entirely \citep{Krumholz07,Bate09,Peters10a,Peters10b,Peters10c}. On the other hand, radiation pressure may indirectly enhance the formation of dense cores and the subsequent condensation into protostars. There are two different scenarios for this triggered star formation process. The first, ''collect and collapse'' \citep{Elmegreen77,Whitworth94}, appears where an expanding \textsc{Hii} region pushes ambient gas into a shell, followed by fragmentation and collapse of the swept-up material. The other one is called ''radiation driven implosion'' \citep{Bertoldi89,Bertoldi90} and depicts the UV-driven compression of a cloud core that is embedded into an \textsc{Hii} region. Apart from radiation, magnetic fields have a noticeable impact on the evolution of a collapsing core. Magnetic pressure and tension forces counter the gravitational collapse and diminish fragmentation without completely inhibiting it \citep{Ziegler05,Banerjee06,Price07,HennebelleTeyssier2008,Hennebelle08,HennebelleCiardi09,Commercon10,Buerzle11,SeifriedEtAl2011}. Likewise, different initial conditions of the gas cloud strongly influence the star formation process \citep[][hereafter Paper~I]{Girichidis11a}.

However, a clear universal formation picture of stellar systems in dense environments is still missing. Several formation and gas accretion processes have been proposed with different distribution scenarios of the gas onto the protostellar objects and different predictions concerning the more massive stars and the resulting IMF. The apparent similarity between the stellar mass function and the mass function of bound cores \citep{Testi98} lead to the monolithic collapse model as a possible star formation scenario, in which every dense protostellar core collapses to a single star \citep{McKee02,McKee03}. However, this scenario lead to a time-scale problem, that can effectively destroy the similarity between IMF and the core mass function \citep{Clark07}. Similar Problems hold for the proposal that massive stars form by collisions of low-mass stars \citep{Zinnecker07}, because the observed stellar densities are too low for this process to be important \citep{Baumgardt11}. As the formation process of a cluster in a dense environment is highly turbulent and dynamic, analytic estimates only give a vague idea of how accretion in this surroundings may work. With the help of simulations, two different cases have been proposed. In one scenario, which is called competitive accretion, the formation of the most massive stars is due to a privileged position close to the centre of the stellar cluster, where the accretion rates are predicted to be highest throughout the simulation \citep{Bonnell01, Bonnell01b}. The other scenario is fragmentation induced starvation \citep{Peters10c}, in which initially the central accretion rates are highest as well. However, subsequent fragmentation shows an increasing impact of the nascent protostars on the accretion behaviour of the central region, that gets starved of material by the surrounding companions.

In this study we investigate the accretion process in different initial density profiles and different turbulent velocity fields. Paper~I focused on the cloud structure and morphology. In this paper we investigate the accretion processes in the formed clusters in detail.

\section{Numerical Methods \& Initial Conditions}

The simulation data used in this study are the same as in Paper~I, where a detailed description of the initial setups can be found. Here we only summarise the key parameters.
 
\subsection{Global Simulation Parameters}
We simulate the collapse of a spherically symmetric molecular cloud with a radius of $R=0.1\,\mathrm{pc}$ and a total mass of $100\,M_\odot$. The resulting average density is $\rhoav = 1.76\times10^{-18}\,\mathrm{g\,cm}^{-3}$ and the corresponding free-fall time gives $t_\mathrm{ff}=5.02\times10^4\,\mathrm{yr}$. The gas with a mean molecular weight of $\mu=2.3$ is assumed to be isothermal at a temperature of $20\,\mathrm{K}$, yielding a constant sound speed of $c_\mathrm{s}=2.68\times10^4\,\mathrm{cm\,s}^{-1}$. The Jeans length, $\lambda_\mathrm{J}$, and the corresponding Jeans mass $M_\mathrm{J}$, calculated as a sphere with diameter $\lambda_\mathrm{J}$, are $\lambda_\mathrm{J}=9300\,\mathrm{AU}$ and $M_\mathrm{J}=1.23\,M_\odot$, respectively. Table~\ref{tab:phys-param} provides an overview of all physical parameters.

\begin{table}
  \caption{Physical parameters of all setups}
  \label{tab:phys-param}
  \begin{tabular}{lcc}
    Parameter & & Value\\
    \hline
    cloud radius & $R$ & $3\times10^{17}\,\mathrm{cm}\approx0.097\,\mathrm{pc}$\\
    total cloud mass & $M_\text{tot}$ &$100\,M_\odot$\\
    mean mass density & $\rhoav$ & $1.76\times10^{-18}$\,g\,cm$^{-3}$\\
    mean number density & $\langle n \rangle$ & $4.60\times10^5$\,cm$^{-3}$\\
    mean molecular weight & $\mu$ & $2.3$\\
    temperature & $T$ & 20\,K\\
    sound speed & $c_\text{s}$ & $0.27$\,km\,s$^{-1}$\\
    rms Mach number & $\mathcal{M}$ & $3.28-3.64$\\
    mean free-fall time & $t_\text{ff}$ & $5.02\times10^4\,\text{yr}$\\
    sound crossing time & $t_\text{sc}$ & $7.10\times10^5\,\text{yr}$\\
    turbulent crossing time & $t_\text{tc}$ & $1.95 - 2.16\times10^5\,\text{yr}$\\
    Jeans length & $\lambda_\text{J}$ & $9.26\times10^3\,\mathrm{AU}\approx0.23\,R_0$\\
    Jeans volume & $V_\text{J}$ & $1.39\times10^{51}\,\text{cm}^3$\\
    Jeans mass & $M_\text{J}$ & $1.23\,M_\odot$\\
    \hline
  \end{tabular}
\end{table}

\subsection{Numerical Code}
The simulations were carried out with the astrophysical code FLASH \citep{FLASH00} in version 2.5. To integrate the hydrodynamic equations, we use a piecewise-parabolic method (PPM) \citep{Colella84}. The code is parallelised using MPI. The computational domain is subdivided into blocks containing a fixed number of cells with an adaptive mesh refinement (AMR) technique based on the PARAMESH library \citep{PARAMESH99}.

\subsection{Resolution and Sink Particles}
The simulations were run with a maximum effective resolution of $4096^3$ grid cells, corresponding to a smallest cell size of $\Delta x\approx13\,\mathrm{AU}$. In order to avoid artificial fragmentation, the Jeans length has to be resolved with at least $4$ grid cells \citep{Truelove97}. At the highest level of refinement, we set the minimum resolution of the Jeans length to $6\Delta x$, which can be transformed to a maximum density of
\begin{equation}
  \rho_\text{max} = \frac{\mathrm{\pi} c_\text{s}^2}{4\,G\,(3\,\Delta x)^2} = 2.46\times10^{-14}\text{g\,cm}^{-3}.
\end{equation}
On all lower levels of refinement, the Jeans length is resolved with at least eight grid cells.
If a cell exceeds this density, a spherical control volume with a radius of $3\Delta x$ is investigated for gravitational collapse indicators. If the collapse criteria \citep{Federrath10a} are fulfilled, an accreting Lagrangian sink particle is formed. While the total mass of the collapsing control volume is of the order of a Jeans mass at that density, the initial mass of the sink particle is only the mass corresponding to the overdensity above the maximum gas density $\rho_\text{max}$. Therefore, the initial mass of the sink particle at the time of formation is typically significantly below the Jeans mass, and can be as low as $10^{-6}\,M_\odot$. Table~\ref{tab:simul-param} lists the simulation and resolution parameters.

\begin{table}
  \caption{Numerical simulation parameters}
  \label{tab:simul-param}
  \begin{tabular}{lcc}
    Parameter & & Value\\
    \hline
    simulation box size & $L_\text{box}$ & $0.26$\,pc\\
    smallest cell size & $\Delta x$ & $13.06$\,AU\\
    Jeans length resolution & & $\ge8~(6^*)$ cells\\
    max. gas density & $\rho_\text{max}$ & $2.46\times10^{-14}$\,g\,cm$^{-3}$\\
    max. number density& $n_\mathrm{max}$ & $6.45\times10^{9}$\,cm$^{-3}$\\
    sink particle accretion radius &$r_\text{accr}$ & $39.17$\,AU\\
    \hline
  \end{tabular}

  \medskip
  $^*$ at highest level of refinement

\end{table}

\subsection{Initial Conditions}
The following four density profiles were used:
\begin{enumerate}
\item Top-hat profile, $\rho=\mathrm{const}$ (TH)
\item Rescaled Bonnor-Ebert sphere. (BE)
\item Power-law profile, $\rho\propto r^{-1.5}$ (PL15)
\item Power-law profile, $\rho\propto r^{-2.0}$ (PL20).
\end{enumerate}
A detailed  description of the profiles can be found in Paper~I.

The turbulence is modelled with an initial random velocity field, originally created in Fourier space, and transformed back into real space. The power spectrum of the modes is given by a power-law function in wavenumber space ($\mathbf{k}$ space) with $E_\text{k}\propto k^{-2}$, corresponding to Burgers turbulence, consistent with the observed spectrum of interstellar turbulence \citep[e.g.,][]{Larson81,Heyer04}. The velocity field is dominated by large-scale modes due to the steep power-law exponent, $-2$, with the largest mode corresponding to the size of the simulation box. Concerning the nature of the $\mathbf{k}$ modes, compressive (curl-free) modes are distinguished from solenoidal (divergence-free) ones. The simulation uses three types of initial fields: purely compressive fields (c), purely solenoidal (s), and a natural (random) mixture (m) of both. The choice of these different turbulent fields was motivated by the strong impact of the nature of the modes on the cloud evolution, found by \citet{Federrath08, Federrath10b}. Notice also, that the scale of the turbulence influences the result significantly \citep{KlessenHeitsch00}, with most realistic results corresponding to large-scale driving adopted here \citep{Klessen01}, where we focus on decaying turbulence with compressive, solenoidal, and mixed modes are considered here.

All setups have supersonic velocities with an rms Mach number
$\mathcal{M} = v_\mathrm{rms}/c_\mathrm{s}$
ranging from $\mathcal{M}=3.28-3.54$ with an average of $\skl{\mathcal{M}}=3.44$.
The sound crossing time through the entire cloud is
$t_\mathrm{sc}(R) = 7.10\times10^5\,\mathrm{yr}$, and
the time for gas with an average velocity of $\skl{\mathcal{M}}c_\mathrm{s}$ to cross the cloud is
$t_\mathrm{tc}(R) = 2.06\times10^5\,\mathrm{yr}$,
respectively.

We do not explicitly impose any rotation on the cloud. Due to the initial turbulence and randomly located velocity patterns in regions with different density, the total net rotational energy does not strictly vanish. However, the ratio $E_\text{rot}/|E_\text{pot}|$ is less than a few times $10^{-3}$, so only very little energy is deposited in global net rotation.

We combine four density profiles with six different turbulent velocity fields (three different compositions of modes with two different random seeds each). Table~\ref{tab:simulation-global-overview} shows a list of all models.
 
\section{Results}

\subsection{Overview}
We follow the cloud collapse until 20\% of the mass is accreted by sink particles. The simulation time, the number of formed protostars, and the mass of the most massive protostar are listed in table~\ref{tab:simulation-global-overview}. A column density plot at the end of each simulation is shown in Paper~I (figures~4 and 5).

\begin{table*}
  \begin{minipage}{\textwidth}
    \caption{List of the runs and their main simulation properties}
    \label{tab:simulation-global-overview}
    \begin{tabular}{lcclccccc}
      $\rho$ & turb. & seed & name & $E_\text{kin}/|E_\text{pot}|$ & $t_\text{sim}$ & $t_\text{sim}/t_\text{ff}$ & $N_\text{sink}$ & $M_\text{mm}$\\
      & mode  & & & & [kyr] & & & $[M_\odot]$\\
      \hline
      TH   & mix & 1 & TH-m-1   & 0.075 &     48.01 & 0.96 & $311$    & $\phn0.86$ \\
      TH   & mix & 2 & TH-m-2   & 0.090 &     45.46 & 0.91 & $429$    & $\phn0.74$ \\
      BE   & com & 1 & BE-c-1   & 0.058 &     27.52 & 0.55 & $305$    & $\phn0.94$ \\
      BE   & com & 2 & BE-c-2   & 0.073 &     27.49 & 0.55 & $331$    & $\phn0.97$ \\
      BE   & mix & 1 & BE-m-1   & 0.053 &     30.05 & 0.60 & $195$    & $\phn1.42$ \\
      BE   & mix & 2 & BE-m-2   & 0.074 &     31.94 & 0.64 & $302$    & $\phn0.54$ \\
      BE   & sol & 1 & BE-s-1   & 0.055 &     30.93 & 0.62 & $234$    & $\phn1.14$ \\
      BE   & sol & 2 & BE-s-2   & 0.074 &     35.86 & 0.72 & $325$    & $\phn0.51$ \\
      PL15 & com & 1 & PL15-c-1 & 0.056 &     25.67 & 0.51 & $194$    & $\phn8.89$ \\
      PL15 & com & 2 & PL15-c-2 & 0.068 &     25.82 & 0.52 & $161$    & $12.3\phn$ \\
      PL15 & mix & 1 & PL15-m-1 & 0.050 &     23.77 & 0.48 & $\phnn1$ & $20.0\phn$ \\
      PL15 & mix & 2 & PL15-m-2 & 0.071 &     31.10 & 0.62 & $308$    & $\phn6.88$ \\
      PL15 & sol & 1 & PL15-s-1 & 0.053 &     24.85 & 0.50 & $\phnn1$ & $20.0\phn$ \\
      PL15 & sol & 2 & PL15-s-2 & 0.069 &     35.96 & 0.72 & $422$    & $\phn4.50$ \\
      PL20 & com & 1 & PL20-c-1 & 0.042 &     10.67 & 0.21 & $\phnn1$ & $20.0\phn$ \\
      \hline
    \end{tabular}
    
    \medskip
    The table shows the setups with their turbulent modes, the random seed for the velocity field and their acronym used in further discussions. The ratio of kinetic to potential energy results from the applied scaling of the turbulent velocity and varies due to the random position of high- and low-velocity regions. The simulation time is given as total time in kyr and in units of the global free-fall time, $t_\text{sim}/t_\text{ff}$. The last two columns show the total number of sink particles $N_\text{sink}$ and most massive particle $M_\text{mm}$, respectively.
  \end{minipage}
\end{table*}

The TH profile needs the longest time to form collapsing over-densities and confine $20\,M_\odot$ in sink particles. During this time of about $45\,\mathrm{kyr}$, the turbulent motions can compress the gas in locally disconnected areas, leading to distinct subclusters of sink particles. The stronger mass concentration in the centre of the BE setups and the resulting shorter collapse and sink particle formation time suppresses the formation of disconnected subclusters in favour of one main central cluster. The corresponding PL15 profiles show very similar cloud structure to the BE runs, but significantly different stellar properties. Due to the much stronger gas concentration in the centre of the cloud, all PL15 setups form one sink particle very early in the simulation. This central sink particle accretes the surrounding gas at a high rate and can grow to a massive protostar before the turbulent motions eventually trigger the formation of collapsing filaments, which produce subsequent sink particles. Due to the stronger mass concentration, the clusters in the PL15 runs are more compact.
In both cases (PL15 and BE), star formation can proceed from inside out. The central region, which is much denser in the PL15 case, forms collapsing filaments earlier. As time proceeds, the turbulence can continuously compress material in outer regions, and thus form sink particles at larger distances from the centre. The PL20 profile only forms one single sink particle due to the very strong mass concentration. It forms very early in the simulation and accretes gas at a constant rate of about $2\times10^{-3}\,M_\odot\,\mathrm{yr}^{-1}$, extremely close to the analytical value of a highly unstable singular isothermal sphere (Paper~I). This results in a total simulation time of only $11\,\mathrm{kyr}$, which is not enough for turbulent motions to form other filaments and further sink particles.

For the analysis of the accretion behaviour in a dense cluster environment, we only consider the setups with multiple sink particles. In oder to understand the accretion process, especially during the early phases of a cluster, one has to distinguish between different star formation scenarios and the resulting cloud morphology around the accreting protostars. Here we discuss two extreme cases, an early disc-like structure around a central protostellar object and an initially filamentary structure that is not dominated by angular momentum. The first case is more pronounced in centrally concentrated density profiles where a net angular momentum with respect to the centre of mass is concentrated in a smaller and denser region. In contrast, initially flatter density structures allow the formation of very massive filaments and subsequently collapsing regions within these filaments out to larger distances from the centre of mass. The angular momentum with respect to the centre of mass may be of the same order as in the former case. However, due to the overall shallower density profile in the centre, the mass infall is slower and the timescale for the formation of a disc in the central region is larger than the timescale for turbulence to form filaments.

\subsection{Limiting Accretion Effects}
Before we discuss the two extreme cluster formation modes and their resulting accretion behaviour, we want to mention two effects that may limit the accretion in a cluster, namely the angular momentum barrier and accretion shielding by companion stars. In order to be accreted onto the surface of a protostar the gas needs to have a specific angular momentum that is lower than the Keplerian angular momentum with respect to the centre of the potential, given by $j_\mathrm{Kepler} = \sqrt{2GM_\mathrm{enc}r\,}$, where $G$ is Newton's constant and $M_\mathrm{enc}$ is the enclosed mass within radius $r$. In a cluster and especially in disc-like environments, angular momentum can quickly be redistributed due to local instabilities, turbulent motions, and strong accretion streams. We analyse the angular momentum of the cloud with respect to the centre of the cluster and calculate the magnitude of the average angular momentum over thin spherical shells.

For the effects of accretion shielding by surrounding companion stars, we estimate how much of the infalling gas can be accreted by secondary cluster members before reaching the centre of the cluster and becoming accreted by the primary cluster stars. Assuming a virialised cluster, one can derive the shielded fraction $S$ using Bondi-Hoyle accretion \citep{Bondi44,Bondi52}. The Bondi-Hoyle radius can be written as
\begin{equation}
  R_\mathrm{BH}=\frac{2GM}{c_\mathrm{s}^2 + v^2},
\end{equation}
where $M$ is the mass of the accreting star, $c_\mathrm{s}$ is the speed of sound, and $v$ is the relative velocity between the star and the immediate surrounding gas. In our analysis we investigate the gas within a radius $r_\mathrm{surr}=100\,\mathrm{AU}$ around each protostar. An application of the Bondi-Hoyle analysis to the accretion in clusters and the cluster dynamics can also be found in \citet{Bonnell01}. The total shielded fraction of the gas falling onto a cluster with $N$ stars, an average stellar mass $M$, and a cluster radius $R_\mathrm{Cl}$ can be estimated by the fraction of the shielded surface area
\begin{equation}
  S= \frac{4\pi\,N\,R_\mathrm{BH}^2}{4\pi R_\mathrm{Cl}^2} = \frac{N}{R_\mathrm{Cl}^2}\rkl{\frac{2GM}{c_\mathrm{s}^2 + v^2}}^2.
\end{equation}
In the case of high relative velocities between the stars and the gas, the $c_\mathrm{s}^2$ term can be neglected. Replacing $v$ with the global virial velocity $v_\mathrm{glob}$ of the cluster,
\begin{equation}
  v_\mathrm{glob} = \rkl{\frac{GM_\mathrm{Cl}}{R_\mathrm{Cl}}}^{1/2} = \rkl{\frac{G\,N\,M}{R_\mathrm{Cl}}}^{1/2},
\end{equation}
with $M_\mathrm{Cl} = N M$ being the cluster mass (assuming equal-mass stars for simplicity), yields
\begin{equation}
  \label{eq:shielding-large-vel}
  S\propto\frac 1 N .
\end{equation}
In that case, the shielding is less efficient the more stellar objects there are in the cluster. Although this might sound counterintuitive, the assumption of virial velocities for the stars explains the strongly reduced Bondi-Hoyle radius with increasing number $N$.
In the other extreme case of no relative velocity of the star to the surrounding gas, the shielding becomes
\begin{equation}
  \label{eq:shielding-small-vel}
  S=\frac{NR_\mathrm{BH}^2}{R_\mathrm{Cl}^2} \propto \frac{NM^2}{c_\mathrm{s}^2R_\mathrm{Cl}},
\end{equation}
linearly increasing with $N$. In our simulations we observe stars forming inside filaments with initially no or very small relative velocity between them and the surrounding gas. The closer to the centre the stars form or migrate during the collapse of the cloud, the more they decouple dynamically from the gas forming a strongly interacting $N$-body system with a gaseous background. We therefore expect the first case ($S\propto1/N$) to be more important in the central region of the cluster, whereas the latter effect ($S\propto N$) is expected to be dominant in the outskirts of the cluster and in the filaments. An analysis of the simulation is shown below, separately for the two extreme formation modes.

\subsection{Disc-like Accretion Mode}
\begin{figure}
  \centering
  \includegraphics[width=8cm]{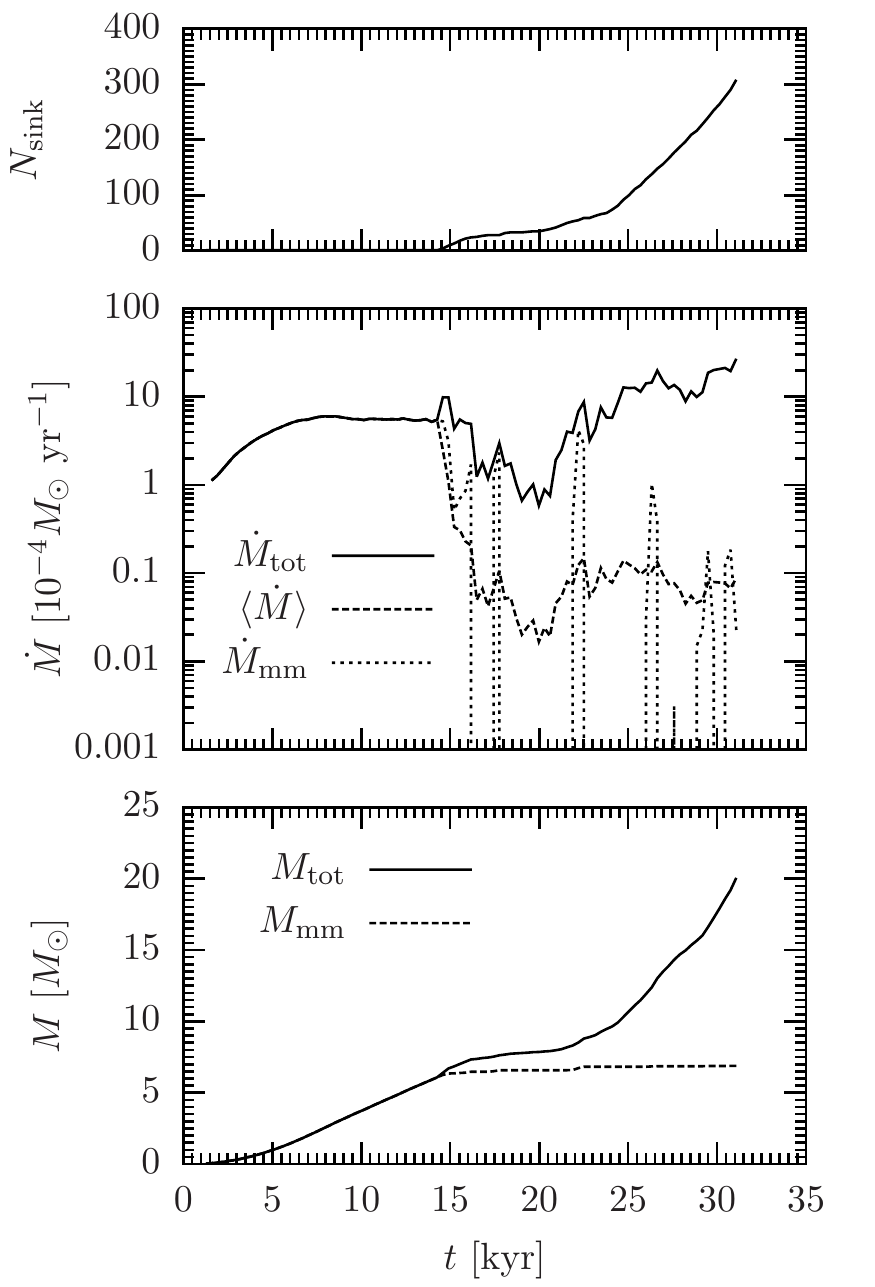}
  \caption{Mass evolution of the most massive, central sink particle in setup PL15-m-2. The upper panel shows the number of sink particles in the setup, the middle panel plots the accretion rate onto the most massive sink particle ($\dot{M}_\mathrm{mm}$) and onto all sink particles ($\dot{M}_\mathrm{tot}$). The bottom panel shows the mass.}
  \label{fig:PL15-m-2-accretion-mm}
\end{figure}

In the first case we discuss the accretion scenario in a disc-like structure, concentrating on simulation PL15-m-2 as an example. A global accretion history for all protostars is shown in figure~\ref{fig:PL15-m-2-accretion-mm}, where we plot the total number of sink particles as well as the accretion rate and the total mass confined in sink particles. The middle panel shows the total accretion rate onto all sink particles $\dot{M}_\mathrm{tot}$, the average accretion rate per protostar $\dot{\skl{M}} = \dot{M}_\mathrm{tot}(t)/N_\mathrm{sink}(t)$, as well as the accretion rate of the most massive protostar $\dot{M}_\mathrm{mm}$, which happens to be the protostar located close to the centre of mass of the gas cloud and the cluster.

\begin{figure}
  \centering
  \includegraphics[width=7.4cm]{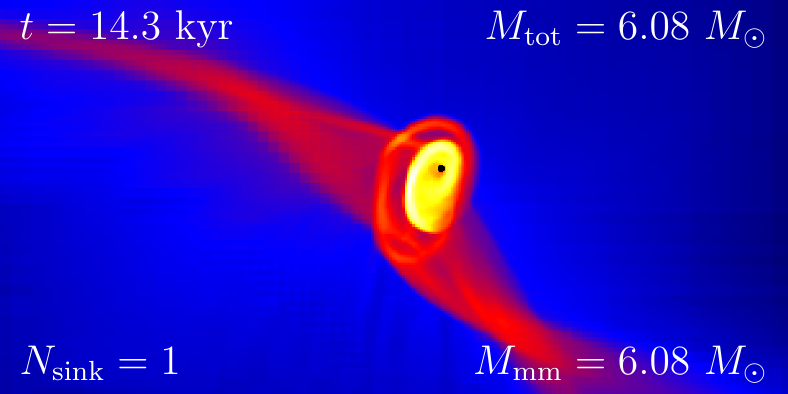}
  \includegraphics[width=7.4cm]{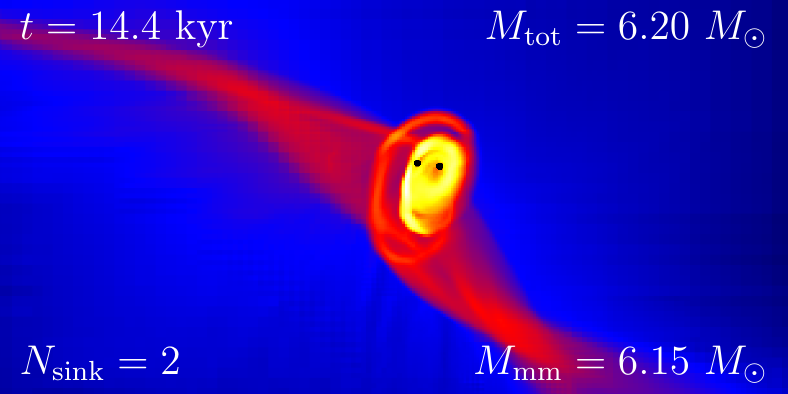}
  \includegraphics[width=7.4cm]{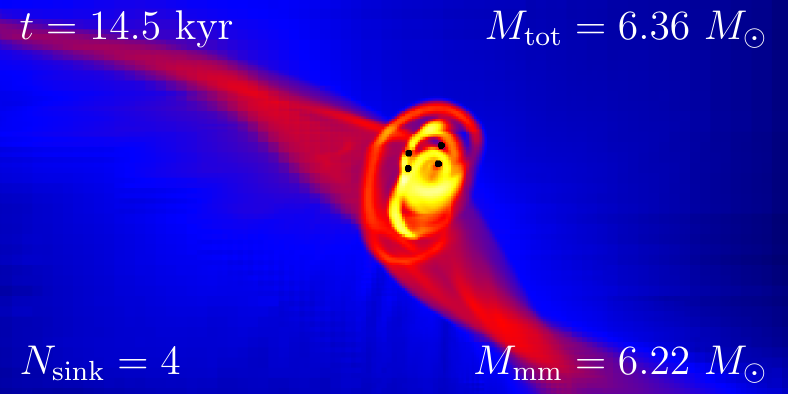}
  \includegraphics[width=7.4cm]{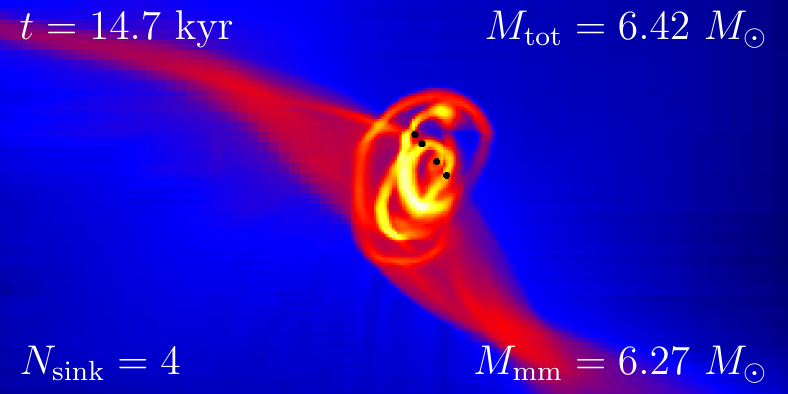}
  \includegraphics[width=7.4cm]{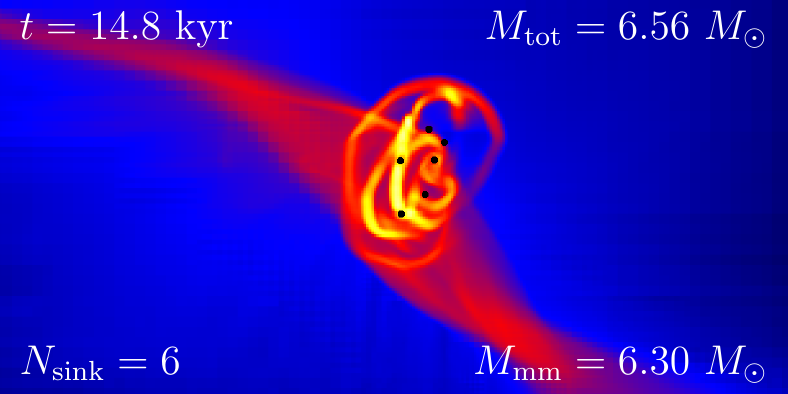}\\
  \ColorBar
  \caption{Column density plots of the central region with the disc around the most massive sink particle (subscript mm) in run PL15-m-2. The formation of secondary sink particles indicates the fragmentation into several objects, which quickly leads to the dissolution of the disc. Spiral arms develop and redirect the gas away from the central protostar, which gets starved of material. The images span roughly $4000\times2000$\,AU.}
  \label{fig:PL15-m-2-coldens-shielding}
\end{figure}

During the first $7\,\mathrm{kyr}$, the turbulent motions do not have enough time to significantly disturb the cloud. The anisotropies formed so far can be neglected and the assumption of a spherically symmetric collapse holds. The first sink particle forms in the centre of the cloud and accretes material in nearly free-fall. The initial density profile results in a linearly increasing accretion rate \citep[see also][]{KlessenHeitsch00,SchmejaKlessen04}. At around $t\sim7\,\mathrm{kyr}$, the turbulent motions form a central filament whose colliding arms concentrate angular momentum in the gas around the central protostar and form a disc-like object. Thus, the spherically symmetric free-fall approximation does not hold any longer. For the time between $7\,\mathrm{kyr}\lesssim t \lesssim14\,\mathrm{kyr}$ the disc is stable and grows in mass and size by accretion from the filamentary arms. During that time, the angular momentum barrier prevents the accretion rate to increase further, resulting in a constant value of $\dot{M}\approx5\times 10^{-4}\,M_\odot\,\mathrm{yr}^{-1}$. At $t\gtrsim14\,\mathrm{kyr}$, the disc becomes unstable, forms spiral arms and fragments into multiple objects. The column density plots in figure~\ref{fig:PL15-m-2-coldens-shielding} show a time sequence of this short period. After the formation of other protostars, the accretion rate onto the central sink particle drops dramatically, while the total accretion rate onto all sink particles increases. The gravitational interactions between the protostars in combination with further infalling gas from the filament disturb the initially disc-like structure and quickly destroy the disc. As a result, a more or less spherically symmetric cluster builds up. The average accretion rate $\dot{\skl{M}}$ decreases by more than an order of magnitude after the formation of secondary stars. The fact that $\dot{M}_\mathrm{mm}$ is on average several orders of magnitude lower than $\dot{\skl{M}}$ indicates that the available gas is not equally shared among the protostars but efficiently shielded from reaching the central region. This indicates that the process of fragmentation induced starvation , first described by \citet{Peters10c} for disc-like structures, also works for more complex geometries. Small amounts of gas that can penetrate through the entire cluster and reach the centre can be seen as episodic accretion spikes.

\begin{figure}
  \centering
  \includegraphics[width=8cm]{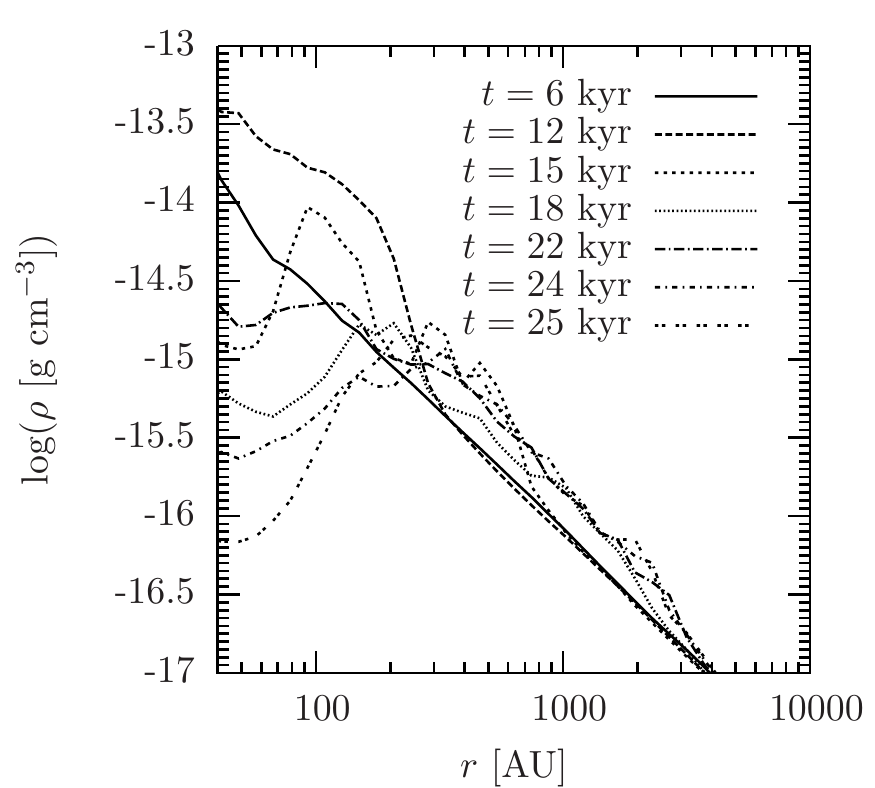}
  \caption{Radial density profile around the central protostar in run PL15-m-2. The gas density first increases in the immediate proximity of the protostar due to the global infall. At later times, the surrounding companions branch off most of the gas, leading to a continuous decrease of the central gas density.}
  \label{fig:PL15m2-cluster-gas}
\end{figure}

\begin{figure}
  \centering
  \includegraphics[width=8cm]{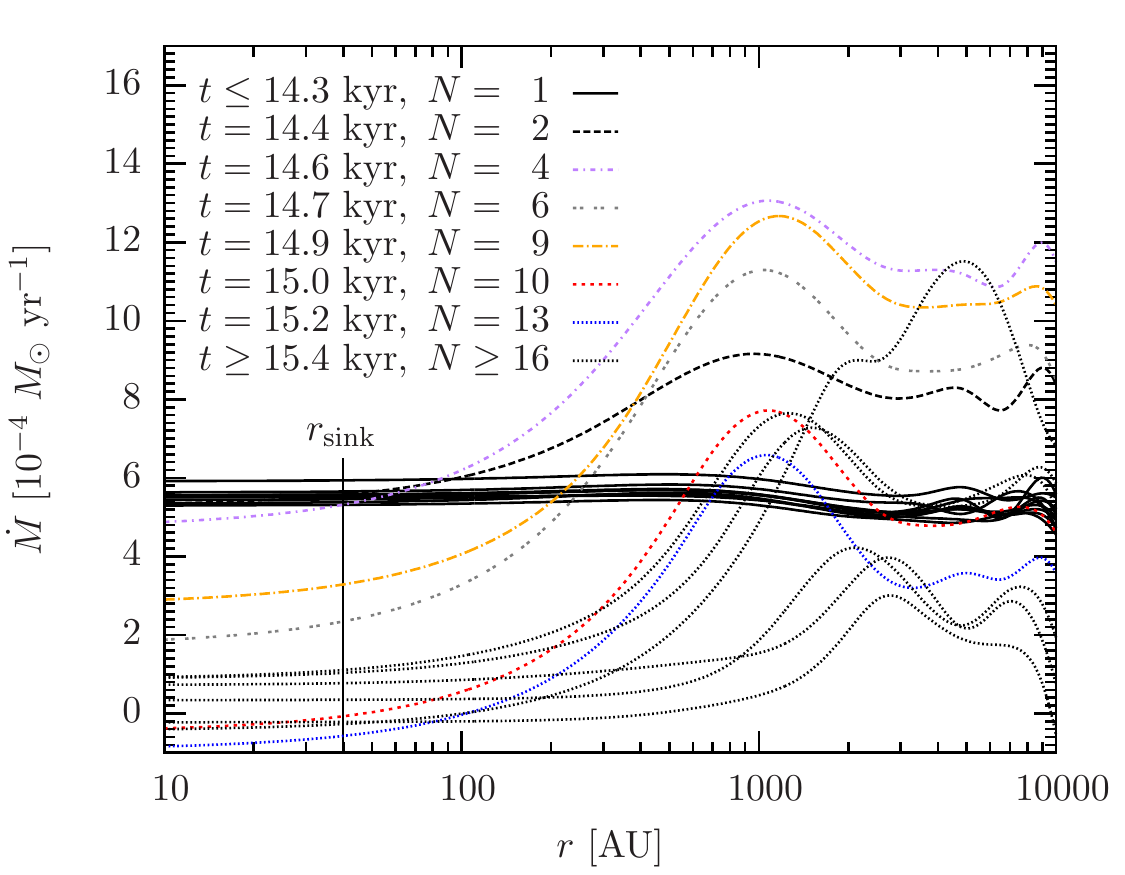}
  \caption{Accretion rate onto spherical shells around the central protostar in the cluster of setup PL15-m-2. The sink particle's accretion radius is indicated by $r_\mathrm{sink}$. The plot shows several curves for $t\le14.3\,\mathrm{kyr}$, where there is only one sink particle ($N=1$). Before the formation of secondary protostars, the accretion rate in the centre is roughly constant. Immediately after surrounding companions have formed, the accretion front moves to larger radii and starves the central object. At later simulation times ($t\ge15.4\,\mathrm{kyr}$) the accretion rate varies, but stays very small for all curves.}
  \label{fig:PL15m2-accr-radius}
\end{figure}

\begin{figure}
  \centering
  \includegraphics[width=8cm]{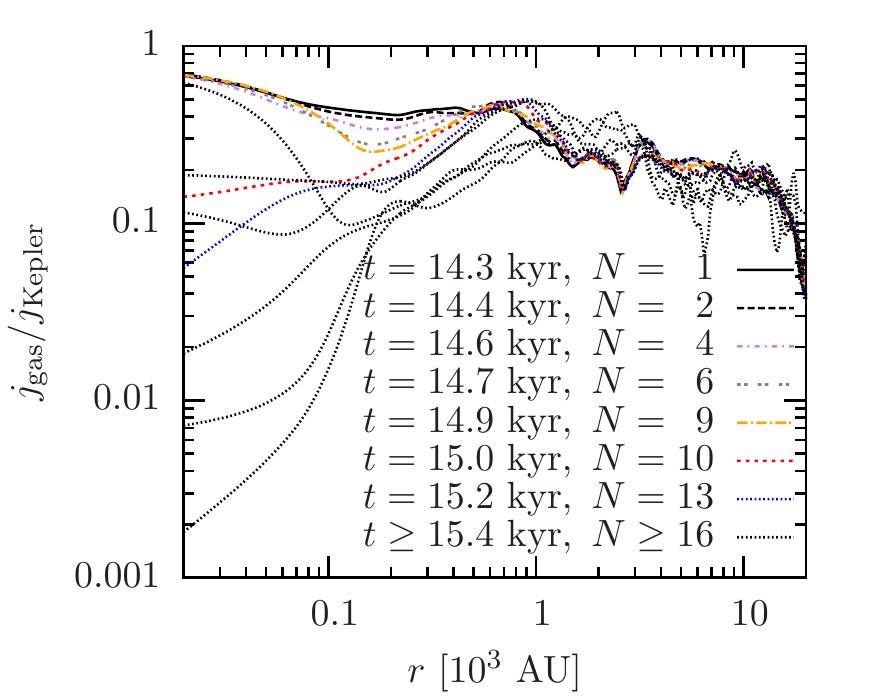}
  \caption{Ratio of specific angular momentum of the gas to the Keplerian value with respect to the centre of the cluster. For all times and all radii the ratio is smaller than unity, indicating that the gas is not prevented from moving inwards by angular momentum. After the formation of subsequent protostars the efficient angular momentum transport even lowers the ratio close to the centre.}
  \label{fig:PL15m2-angmom}
\end{figure}

We emphasise the influence of different geometrical shapes of the cluster. Whereas in \citet{Peters10c} the angular momentum vector was well defined and the starvation effect was complete, the turbulent motions in our simulation allow for some accretion channels. The starvation effect can be understood by plotting the gas density around the central protostar (figure~\ref{fig:PL15m2-cluster-gas}) and the accretion rate onto spherical shells around the central protostar (figure~\ref{fig:PL15m2-accr-radius}), including the net accretion onto gas shells as well as the accretion onto other sink particles. The gas density within a radius of $r\sim300\,\mathrm{AU}$ around the central protostar first increases due to the global infall, forming a dense disc. Once other companions have formed, it decreases continuously, which starves the central object. The accretion rate as a function of radius supports this starvation picture. Before subsequent sink particles form, the accretion rate is roughly constant ($\dot{M}\approx5\times 10^{-4}\,M_\odot\,\mathrm{yr}^{-1}$). Immediately after the disc fragments, the accretion front moves outwards to larger radii, resulting in significantly smaller values around the central protostar. As the protostars in the cluster as well as the gas undergo strong dynamical interactions, the net accretion rate exhibits local fluctuations. The curves for $t\ge15.4\,\mathrm{kyr}$ are therefore averaged over several thousand years. Phases that include an accretion spike (see second panel in figure~\ref{fig:PL15-m-2-coldens-shielding}) lead to positive values at $r_\mathrm{sink}$, at other times the net accretion is negative, because some gas that is not bound to a protostar can enter and exit a spherical shell without being accreted onto any protostar. During the entire simulation time, but in particular after the formation of subsequent protostars, the specific angular momentum in the disc-like object (see figure~\ref{fig:PL15-m-2-coldens-shielding}) as well as in the later formed spherically symmetric cluster is lower than the Keplerian value for all radii (figure~\ref{fig:PL15m2-angmom}). Initially the ratio is close to unity indicating that gas motion in the disc-like object is affected by the angular momentum. This also leads to a deviation from the free-fall accretion rate for the simulation time $t\in(7-14)\,\mathrm{kyr}$, which would further increase in case of no angular momentum. After the disc becomes unstable and further protostars form, the angular momentum is transfered efficiently resulting in a decreasing ratio. The low central accretion rates can therefore not be explained by an angular momentum barrier that inhibits the gas flow to smaller radii. A plot of the angular momentum of the gas as a function of enclosed mass for different times is shown in the appendix (figure~\ref{fig:PL15m2-angmom-Menc}).

\begin{figure}
  \centering
  \includegraphics[width=8cm]{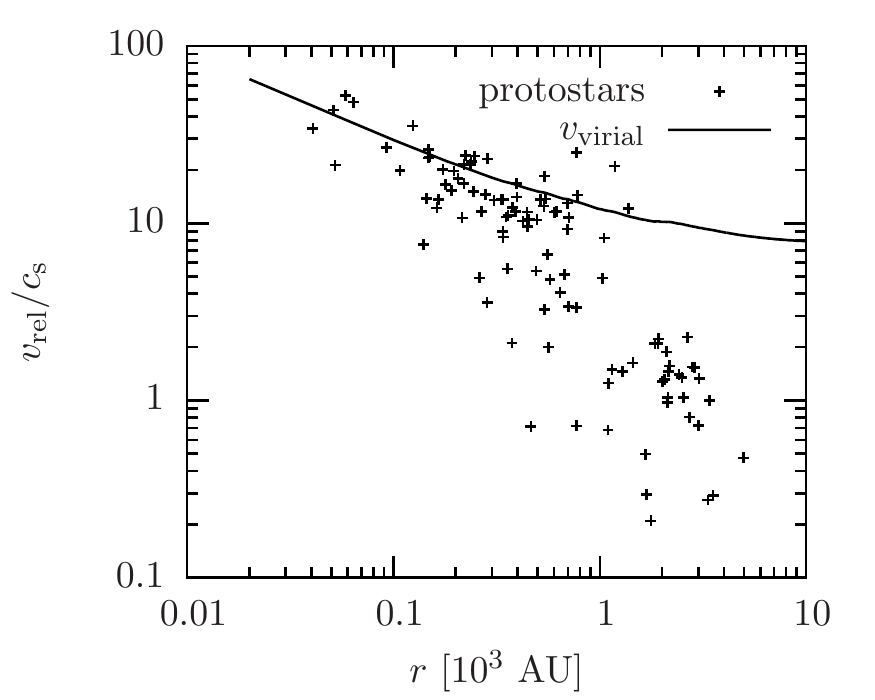}
  \caption{Relative velocity between the protostars and the surrounding gas within a radius of $r_\mathrm{surr}=100\,\mathrm{AU}$ around every protostar as a function of distance from the centre of the cluster $r$ for one snapshot at $t=25\,\mathrm{kyr}$. The plots for different times look very similar. The relative velocity decreases for larger radii ranging from super-virial velocities in the centre down to sub-virial velocities of the order of the sound speed at large distances from the centre.}
  \label{fig:PL15m2-relvel}
\end{figure}

\begin{figure}
  \centering
  \includegraphics[width=8cm]{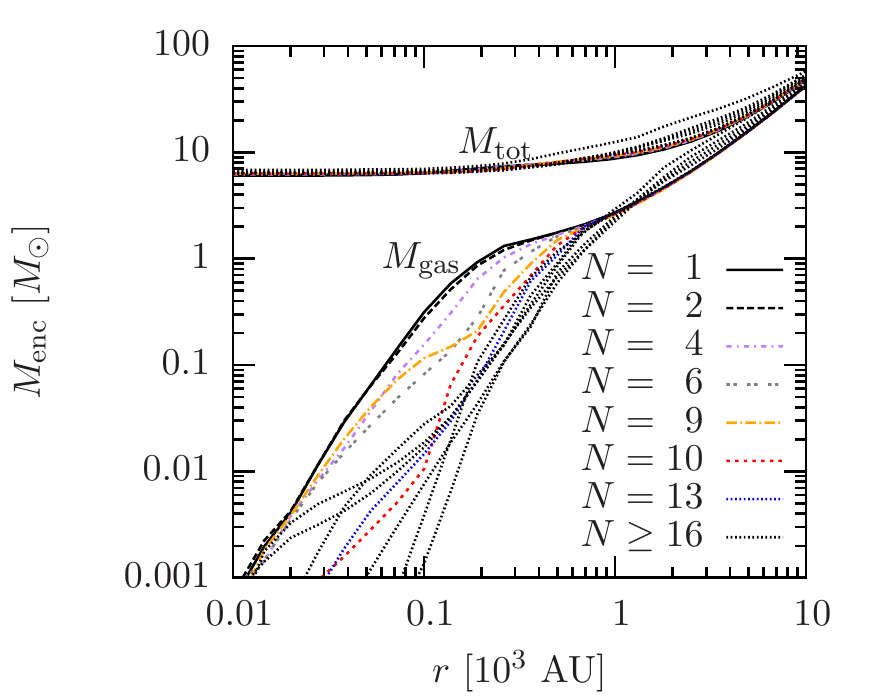}
  \caption{Enclosed masses as a function of radius for the total mass (gas and protostars) and the gas mass. The gas is only a small fraction of the total mass in the centre and decreases strongly after the formation of multiple protostars.}
  \label{fig:PL15m2-rad-masses}
\end{figure}

In addition we analyse which accretion shielding is dominant in different regions of the cloud. In figure~\ref{fig:PL15m2-relvel} we plot the relative velocity of the protostars to the surrounding gas within a radius of $r_\mathrm{surr} = 100\,\mathrm{AU} = 2.5\,r_\mathrm{accr}$ around the protostar. For all times the velocities range from super-virial velocities in the centre of the cloud down to small relative velocities of the order of the speed of sound in the outskirts, indicating that the accretion shielding in the centre follows equation~(\ref{eq:shielding-large-vel}), whereas in the outskirts, it is better described by equation~(\ref{eq:shielding-small-vel}). As expected, both cases of the accretion shielding effects apply. In the central region, the shielding therefore becomes less efficient with an increasing number of protostars. In contrast, the protostars in the outskirts of the cluster are much more efficient in shielding the accretion flow. In addition to the larger shielded fraction, the shielding at larger distances from the centre becomes even more efficient, because the protostars form along the dense filaments that channel the accretion streams. The area of the strongest accretion flow then coincides with the position of the protostars that move with the gas flow. As a result, the total gas mass in the centre of the cluster is very small. Figure~\ref{fig:PL15m2-rad-masses} shows the total enclosed mass (gas and protostars) as well as the enclosed gas mass as a function of radius. The gas mass is only a small fraction of the total mass and decreases strongly in the centre after the formation of multiple objects. Therefore the central accretion rate is low in spite of a less efficient accretion shielding. The gas is already accreted further out before it reaches the less shielded region.

Depending on the definition of the centre of the cluster (position of the most massive protostar, centre of mass of the protostars, centre of mass including protostars and gas), the most massive protostar is not located exactly at the centre of the cluster. In addition, the interaction with other protostars leads to small displacements during the simulation. Neglecting the fact, that only little mass is left in the centre of the cluster, it could be possible that the low accretion rates onto the protostars in the central region arise from these displacements, if the protostar escapes form a collimated accretion stream. In order to verify, that the central stars are really shielded, we need to analyse the motions of the central stars relative to the surrounding gas. We calculate the relative velocity dispersion of the gas with respect to the most massive protostar within a radius of $1000\,\mathrm{AU}$ around it, covering the vast majority of the entire cluster. We find velocity dispersions ranging from $\sigma/c_\mathrm{s}=5.4-10.6$, resulting in a total time for the gas to reach the central star of $1.7-3.3\,\mathrm{kyr}$ after entering the sphere of $1000\,\mathrm{AU}$. This time is only $5-10\%$ of the simulated time. Given the low angular momentum in the centre of the cluster (figure~\ref{fig:PL15m2-angmom}), the available gas has enough time to reach the most massive star regardless of displacements.

\subsection{Filamentary Accretion Mode}
\begin{figure}
  \centering
  \includegraphics[width=8cm]{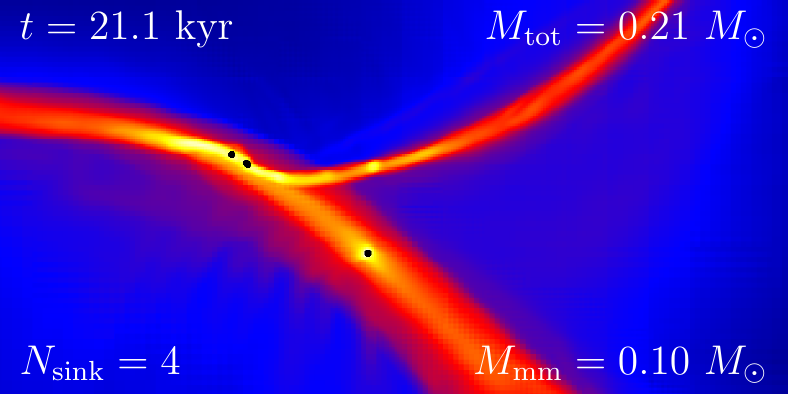}
  \includegraphics[width=8cm]{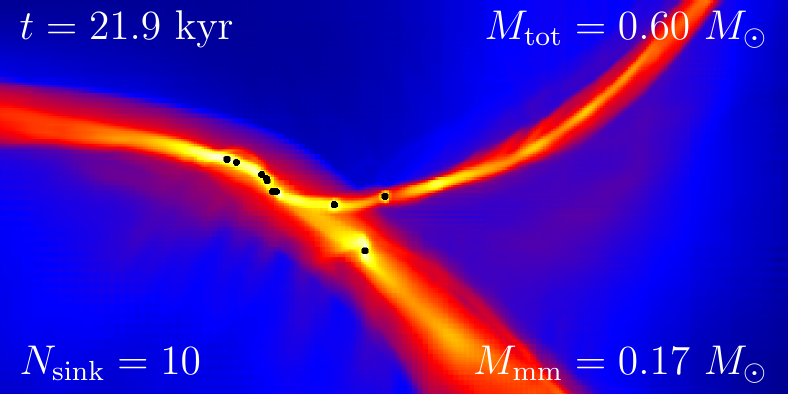}
  \includegraphics[width=8cm]{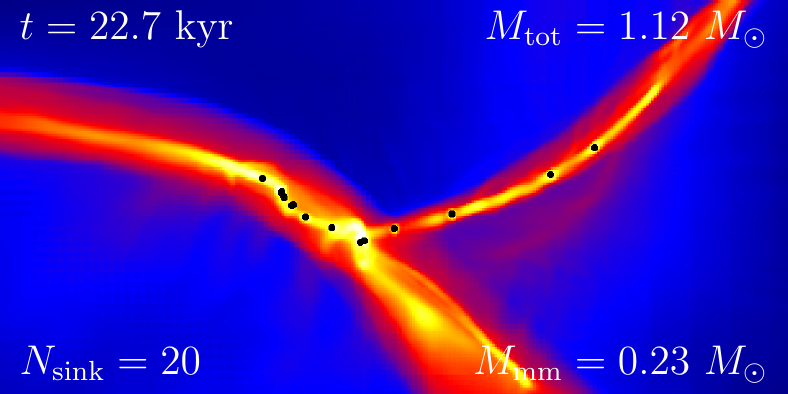}
  \includegraphics[width=8cm]{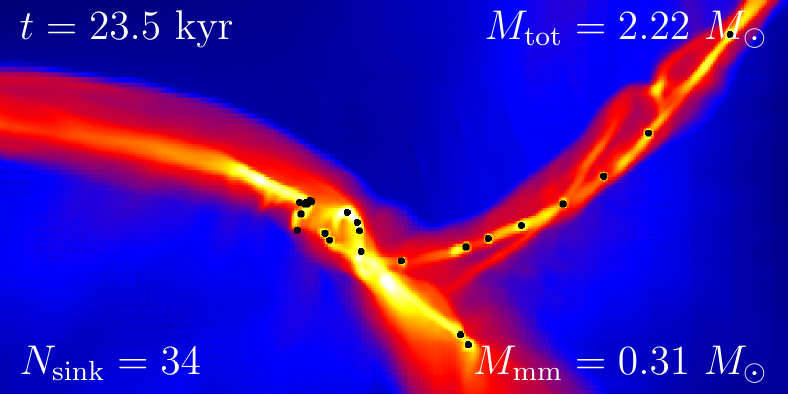}\\
  \ColorBar
  \caption{Column density plots of the central filament in BE-m-2. The sink particles form in the filament and remain there while converging to the centre of the cluster. The closer the protostars approach each other, the stronger decoupled is their motion from the motion of the filament. The images span roughly $4000\times2000$\,AU.}
  \label{fig:BE-m-2-coldens-shielding}
\end{figure}

\begin{figure}
  \centering
  \includegraphics[width=8cm]{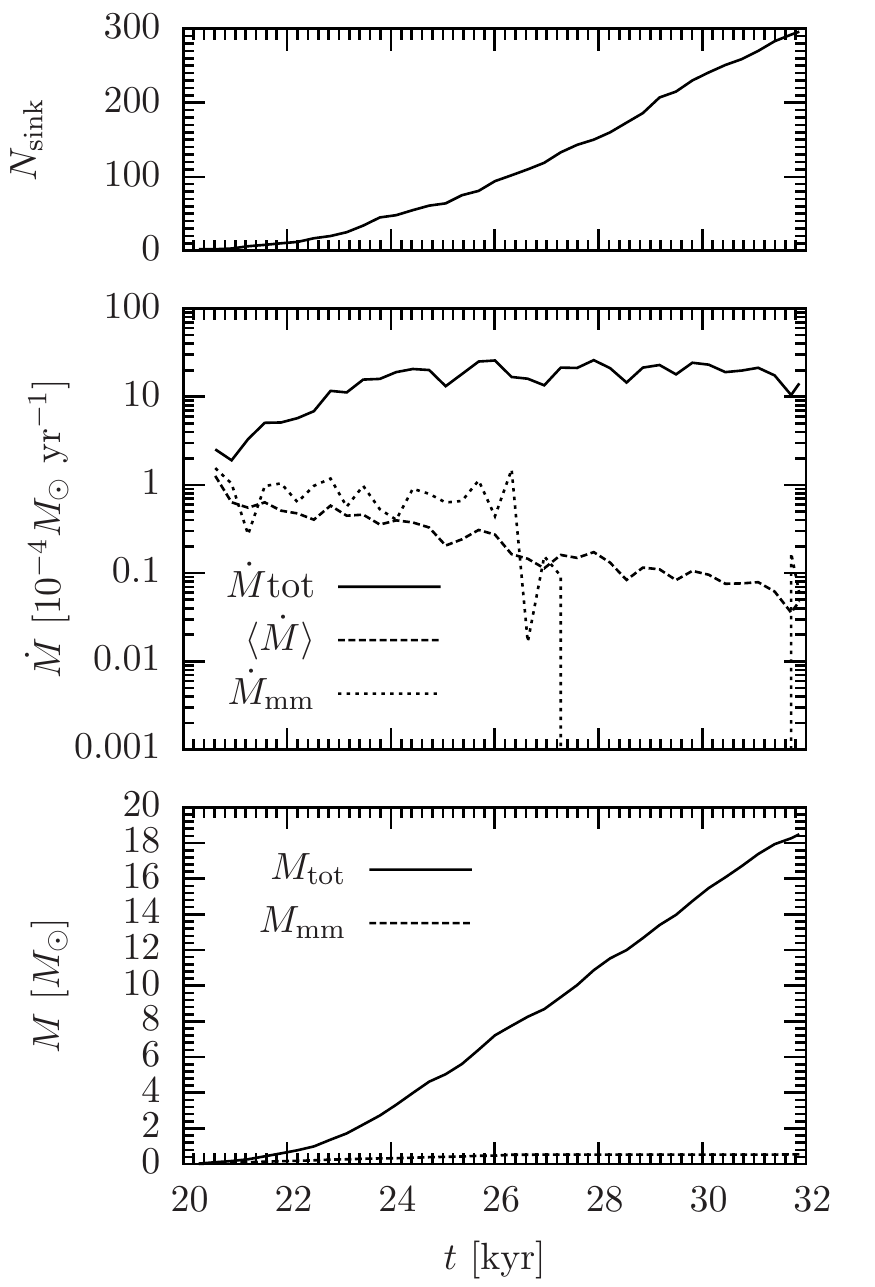}
  \caption{Same as figure \ref{fig:PL15-m-2-accretion-mm} but for the simulation BE-m-2 (filamentary accretion).}
  \label{fig:BE-m-2-accretion-mm}
\end{figure}

If the formation of a cluster is not dominated by a disc-like structure but by filaments, the formation scenario of protostars is different \citep[see, e.g.,][]{Banerjee06}. The initial density profiles with a flat core need more time to form Jeans-unstable regions. During this time, the turbulence can form filaments in which protostellar condensations form next to each other along the densest part of the filament. The column density plots in figure~\ref{fig:BE-m-2-coldens-shielding} show a time evolution of the filamentary collapse. During their formation, the protostars inherit the motion of the parental filament and move with the gas flow. With increasing proximity to other protostars, their attraction as an $N$-body system becomes stronger than the force between protostars and gas. The protostars then dynamically decouple from the filaments and accumulate in the central region in a more spherically-symmetric configuration rather than in a flat or string-like structure. The initial filaments get dispersed in the central region, because the $N$-body system efficiently stirs the gas. The formation of the first protostars is shown in the column density plots in figure~\ref{fig:BE-m-2-coldens-shielding}. The mass accretion in this formation mode is plotted in figure~\ref{fig:BE-m-2-accretion-mm}. Note that figure~\ref{fig:BE-m-2-coldens-shielding} only covers a small time range at the beginning of the cluster formation, whereas figure~\ref{fig:BE-m-2-accretion-mm} covers the entire simulation time after the formation of sink particles.

\begin{figure}
  \centering
  \includegraphics[width=8cm]{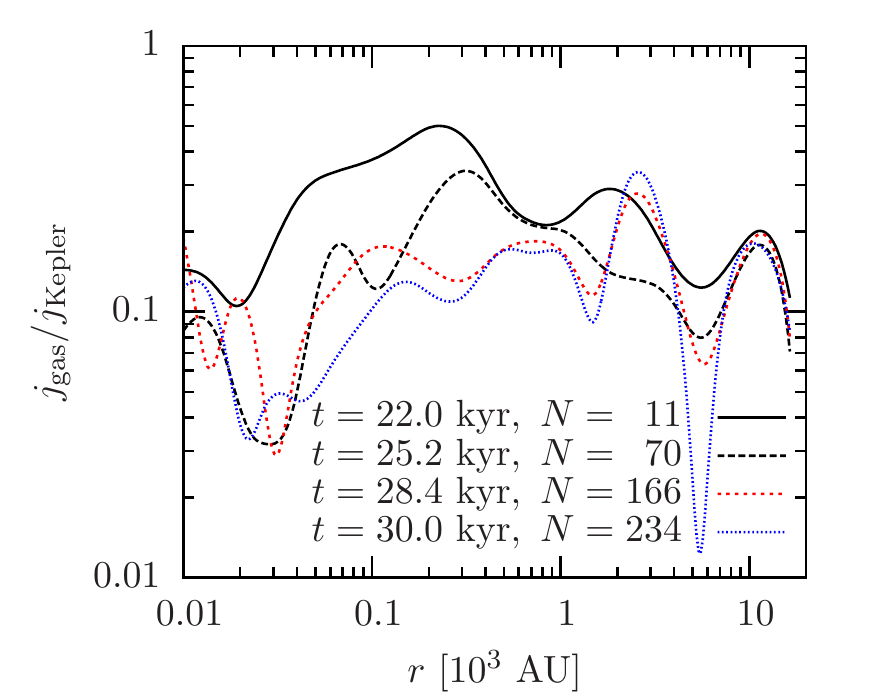}
  \caption{Same as figure~\ref{fig:PL15m2-angmom} but for setup BE-m-2 (filamentary accretion).}
  \label{fig:BEm2-angmom}
\end{figure}

During the first $\sim6\,\mathrm{kyr}$ after the first protostar has formed, the accretion rate onto the most massive protostar stays roughly constant. About 100 sink particles form along the main filaments within this time. Initially, most protostars can accrete gas from both sides of the filamentary arm, resulting in a high accretion rate. Despite the fact that the total accretion rate increases by a factor of $\sim10$, the average accretion rate gradually drops by about one order of magnitude, because this mass flow is distributed between $\sim100$ protostars. This is in contrast to the setup PL15-m-2, where the global accretion rate suddenly drops. As the setup is less concentrated, the Keplerian specific angular momentum of the gas with respect to the centre of the nascent cluster is significantly lower than in the PL15-m-2 case. Nonetheless, the ratio $j_\mathrm{gas}/j_\mathrm{Kepler}$ (figure~\ref{fig:PL15m2-angmom}) is even smaller than in the disc-like accretion mode (figure~\ref{fig:BEm2-angmom}), indicating that the gas motion is not restricted by the angular momentum barrier. A plot of the angular momentum of the gas as a function of enclosed mass for different times is shown in the appendix (figure~\ref{fig:BEm2-angmom-Menc}).

\begin{figure}
  \centering
  \includegraphics[width=8cm]{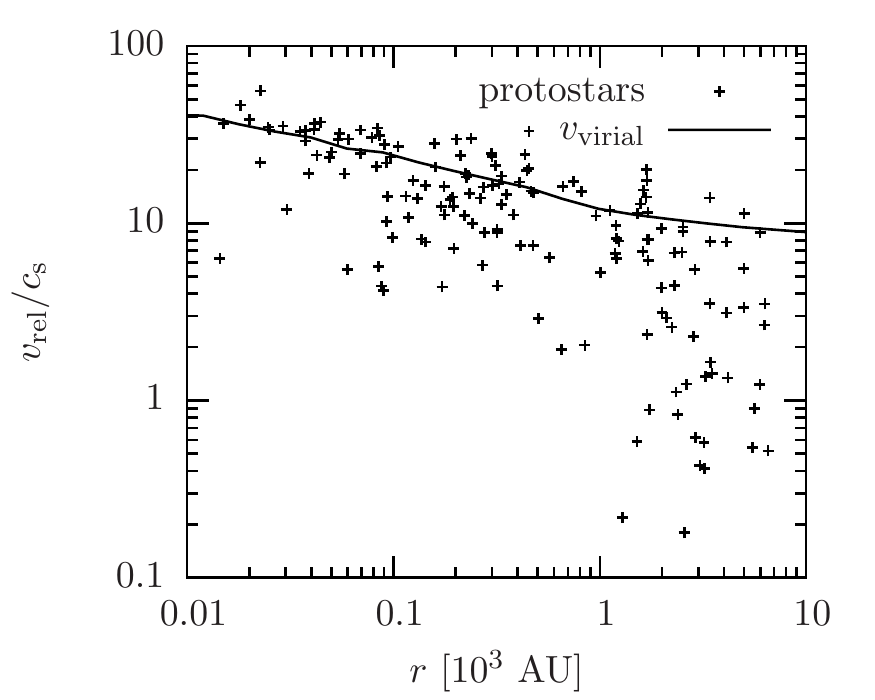}
  \caption{Same as figure~\ref{fig:PL15m2-relvel} but for setup BE-m-2 (filamentary accretion) at $t=28\,\mathrm{kyr}$.}
  \label{fig:BEm2-relvel}
\end{figure}

\begin{figure}
  \centering
  \includegraphics[width=8cm]{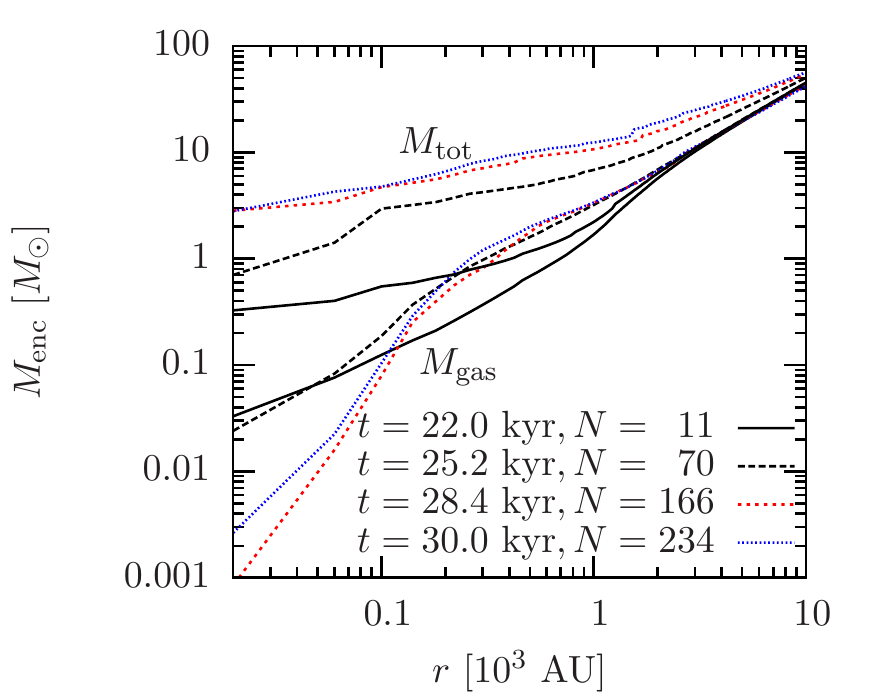}
  \caption{Same as figure~\ref{fig:PL15m2-rad-masses} but for BE-m-2 (filamentary accretion).}
  \label{fig:BEm2-rad-masses}
\end{figure}

The lower central mass concentration allows for extended filaments and the formation of protostars at larger distances from the centre of the cluster. The relative velocity between the protostars and the surrounding gas (figure~\ref{fig:BEm2-relvel}) drops significantly for larger radii, but is closer to the virial velocity for this setup in comparison to the disc-like structure. The global cluster dynamics should therefore correspond to the shielding relation in equation~(\ref{eq:shielding-large-vel}). However, the relative velocities in the outskirts of the cluster are of the order of the speed of sound which marks the transition between the two extreme shielding cases. In addition, the gas distribution in the filamentary structure strongly deviates from spherical symmetry. The protostars in the outskirts of the nascent cluster move along the densest part of the filament and can therefore efficiently accrete a significant fraction of the filament mass before they dynamically decouple from the filament. As a consequence, the accretion shielding is more efficient, the more protostars form along the filaments.
Similar to the disc-like case, the gas content in comparison to the total mass is very low in the central region of the cluster (figure~\ref{fig:BEm2-rad-masses}). Overall, the more protostars accumulate in the central region forming a cluster, the more efficient is the accretion shielding effect. In the case of the most massive sink particle in BE-m-2, accretion is entirely shut off for $t\gtrsim27\,\mathrm{kyr}$.

\subsection{Mass Evolution of all Runs}
\begin{figure}
  \centering
  \includegraphics[width=8cm]{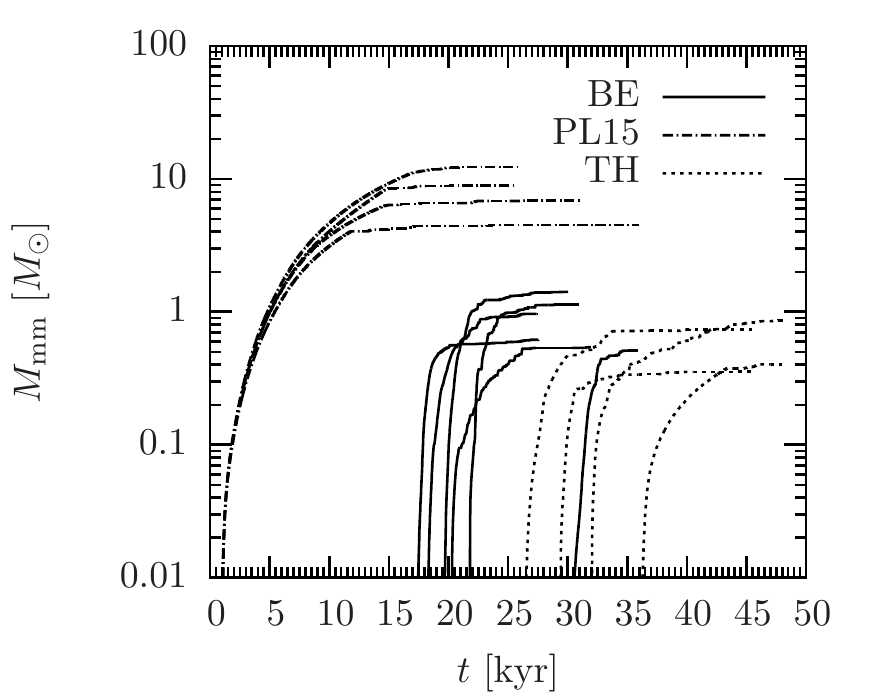}
  \caption{Mass evolution of the most massive protostar for all simulated clusters. In all cases the mass curves flatten significantly at the late-time evolution of the simulation. For a detailed mass analysis of each setup, see figures~\ref{fig:TH-mass-evol-single-sp}--\ref{fig:PL15-mass-evol-single-sp}.}
  \label{fig:MM-combined-masses}
\end{figure}

\begin{figure*}
  \begin{minipage}{\textwidth}
    \centering
    \includegraphics[width=5.5cm]{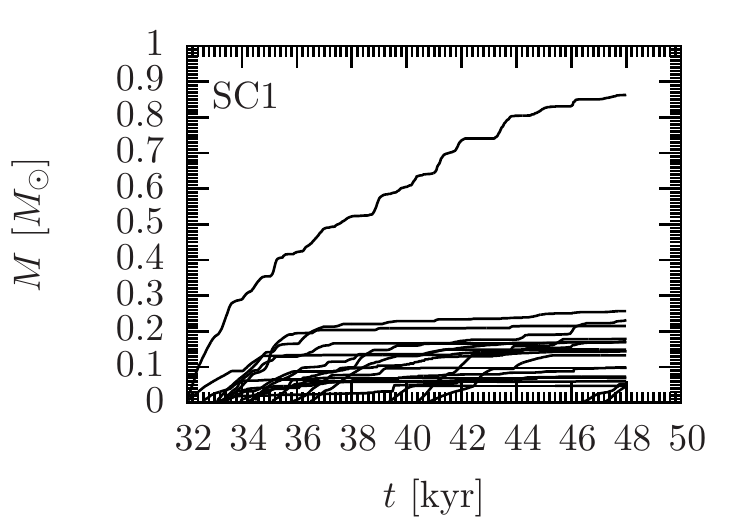}
    \includegraphics[width=5.5cm]{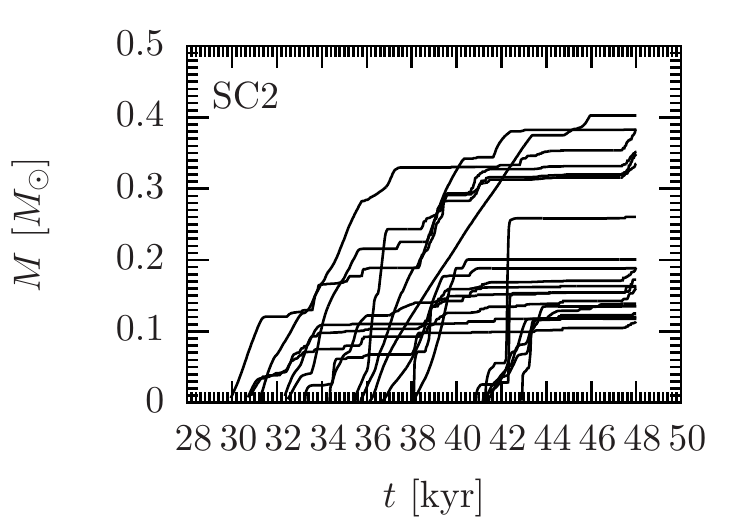}\\
    \includegraphics[width=5.5cm]{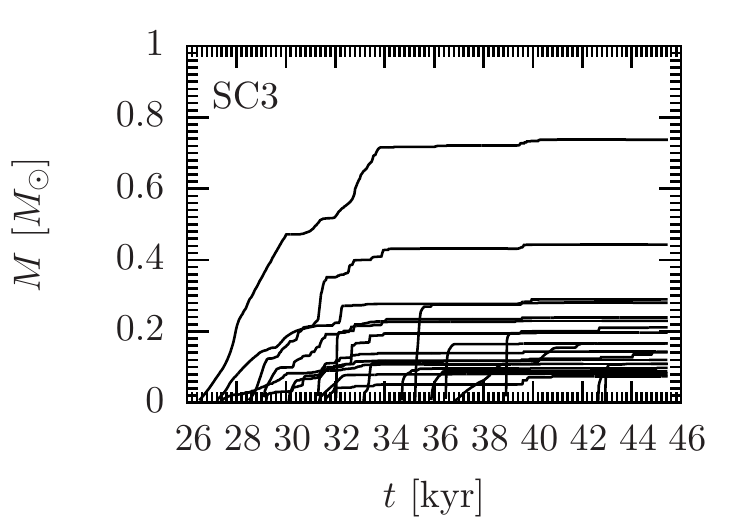}
    \includegraphics[width=5.5cm]{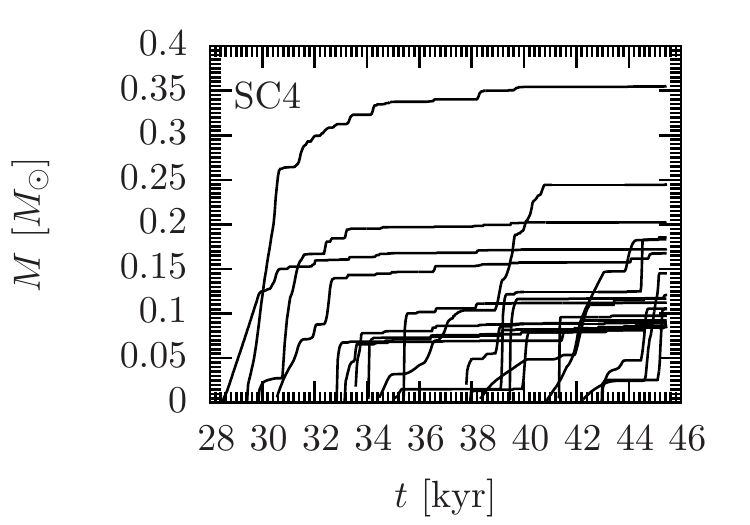}
  \end{minipage}
  \caption{Mass evolution of each sink particle for the four main subclusters in the TH setups. The upper two plots correspond to the biggest subclusters (SC1, SC2) in run TH-m-1. The lower plots (SC3, SC4) are the main subclusters in TH-m-2, respectively. The most massive particles are located closer to the centre of mass and thus experience an efficient starvation effect, which can be seen in the low increase in mass.}
  \label{fig:TH-mass-evol-single-sp}
\end{figure*}

\begin{figure*}
  \begin{minipage}{\textwidth}
    \centering
    \includegraphics[width=5.5cm]{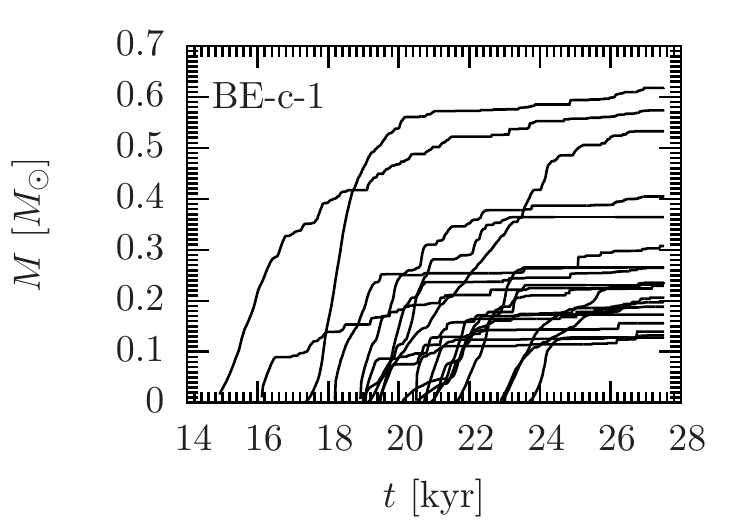}
    \includegraphics[width=5.5cm]{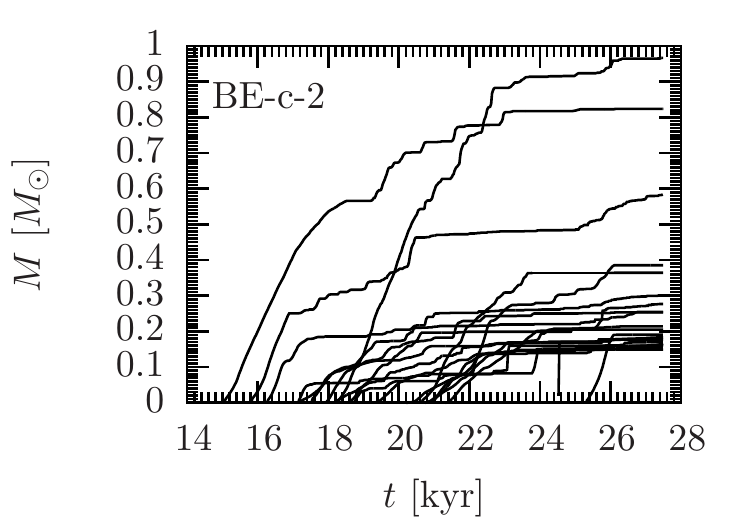}
    \includegraphics[width=5.5cm]{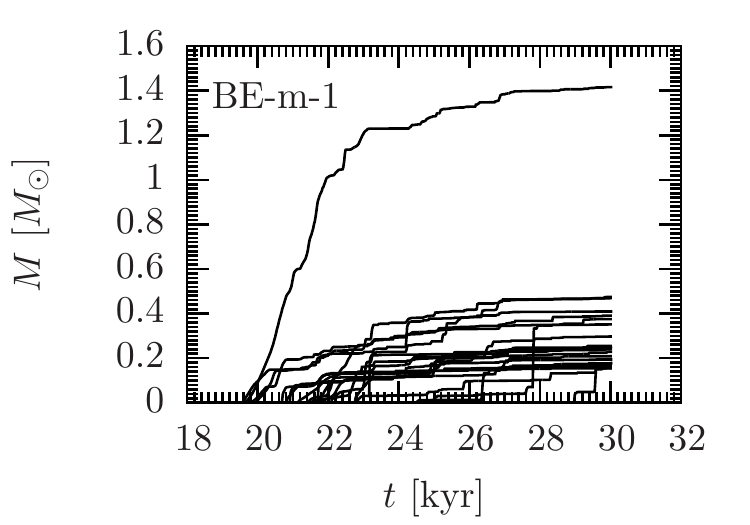}\\
    \includegraphics[width=5.5cm]{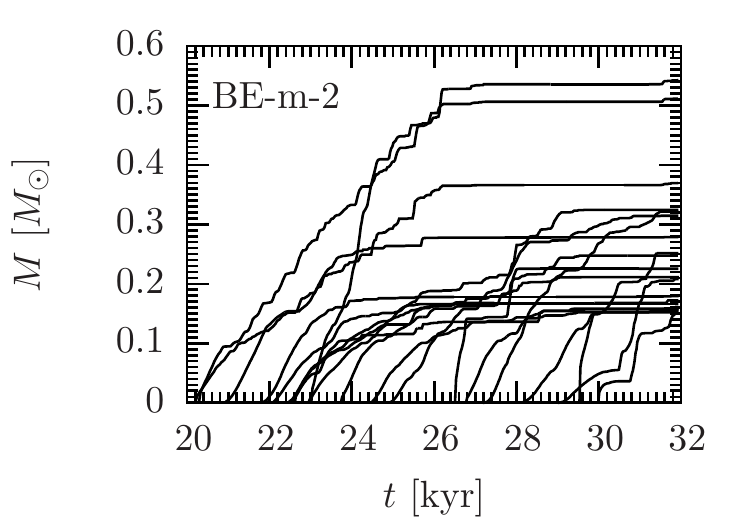}
    \includegraphics[width=5.5cm]{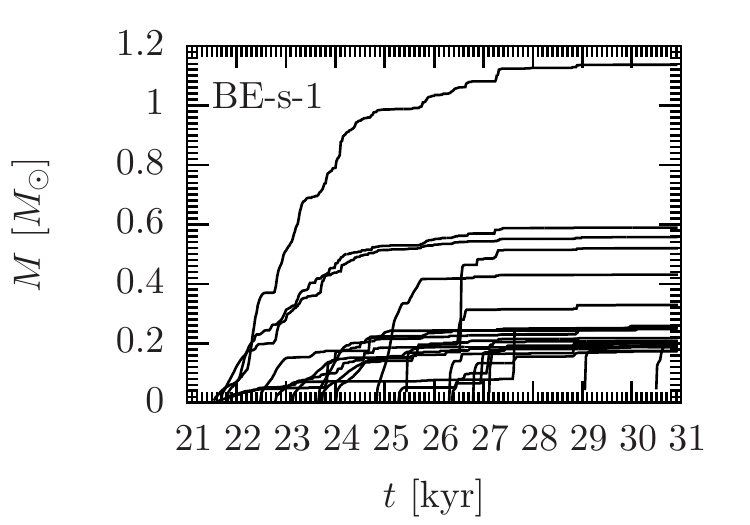}
    \includegraphics[width=5.5cm]{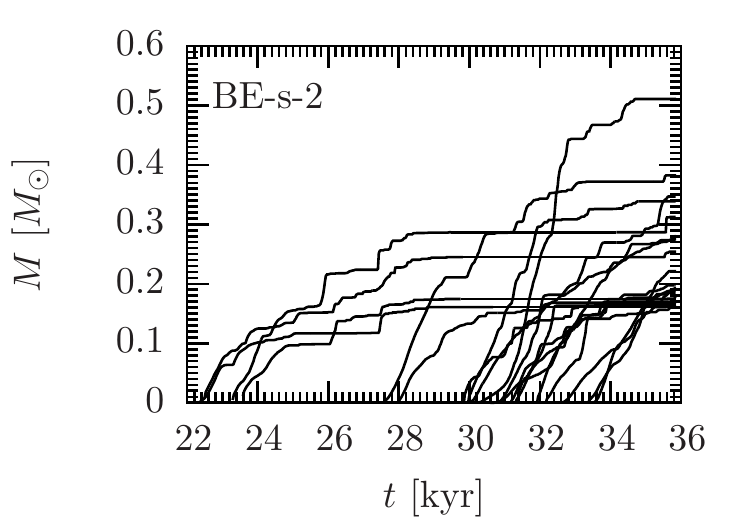}
  \end{minipage}
  \caption{Same as figure~\ref{fig:TH-mass-evol-single-sp} but for the BE setups.}
  \label{fig:BE-mass-evol-single-sp}
\end{figure*}

\begin{figure*}
  \begin{minipage}{\textwidth}
    \centering
    \includegraphics[width=5.5cm]{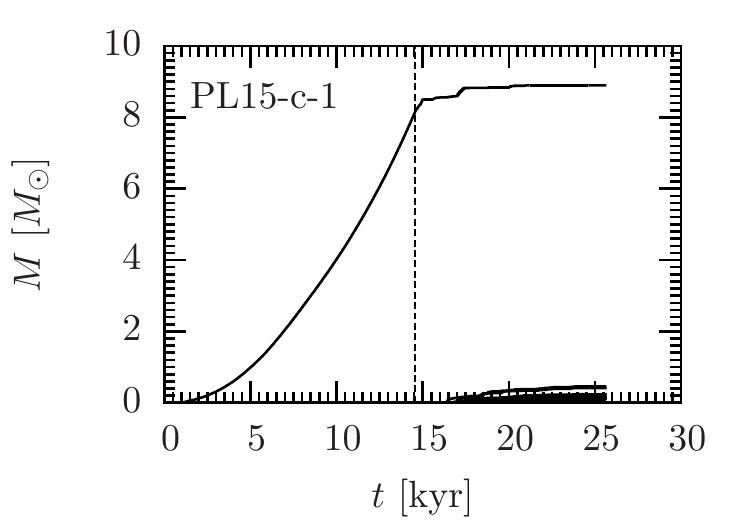}
    \includegraphics[width=5.5cm]{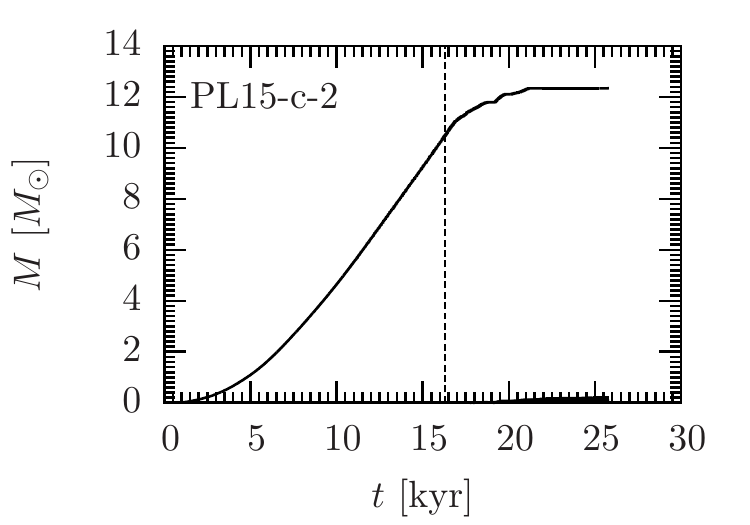}\\
    \includegraphics[width=5.5cm]{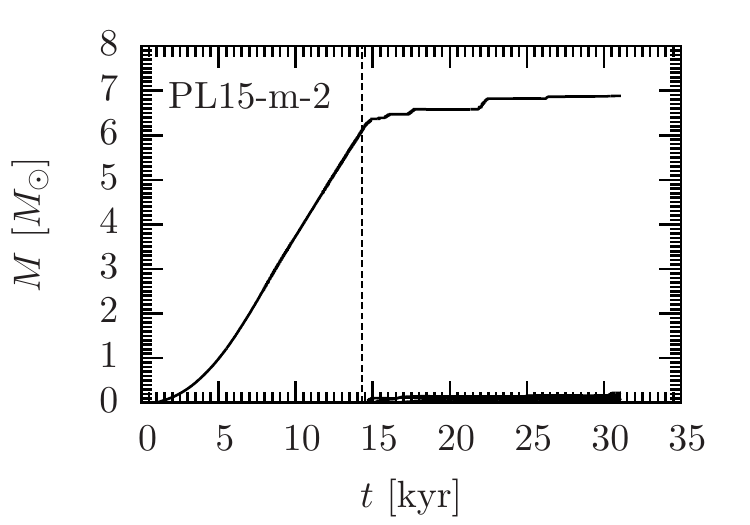}
    \includegraphics[width=5.5cm]{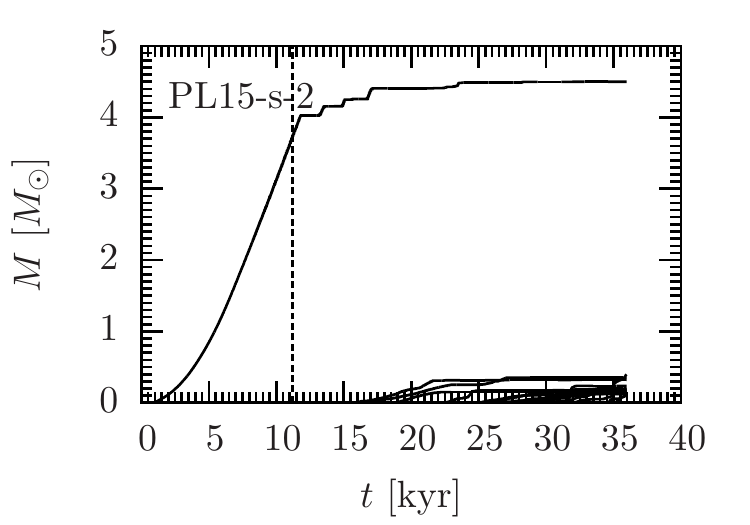}
  \end{minipage}
  \caption{Same as figure~\ref{fig:TH-mass-evol-single-sp} but for the PL15 setups. Note that the masses are plotted in linear scale in order to better see the starvation effect, which starts after the formation of the second sink particle (vertical lines).}
  \label{fig:PL15-mass-evol-single-sp}
\end{figure*}

In general, the formation scenario of protostars in dense clusters will be a mixture of star formation in a disc and filaments, where the above examples are extremes. However, in all simulated cases, the formation of multiple protostars finally leads to a shielding effect in the formed cluster. Figure~\ref{fig:MM-combined-masses} shows the mass evolution of the most massive protostars of all clusters combined, figures~\ref{fig:TH-mass-evol-single-sp}--\ref{fig:PL15-mass-evol-single-sp} show the mass evolution of the 20 most massive sink particles for each simulations. In the case of an initially uniform density (TH, figure~\ref{fig:TH-mass-evol-single-sp}), the two main subclusters for each run (TH-m-1 and TH-m-2) were evaluated separately. The corresponding accretion plots for the most massive sink particle are very similar to the ones shown in figures~\ref{fig:PL15-m-2-accretion-mm} and \ref{fig:BE-m-2-accretion-mm}. Despite different formation scenarios, all setups show very similar structures and emphasise the starvation effect on the most massive protostars, which are located preferentially closer to the centre of the cluster than low-mass companions.

\begin{figure}
  \centering
  \includegraphics[width=8cm]{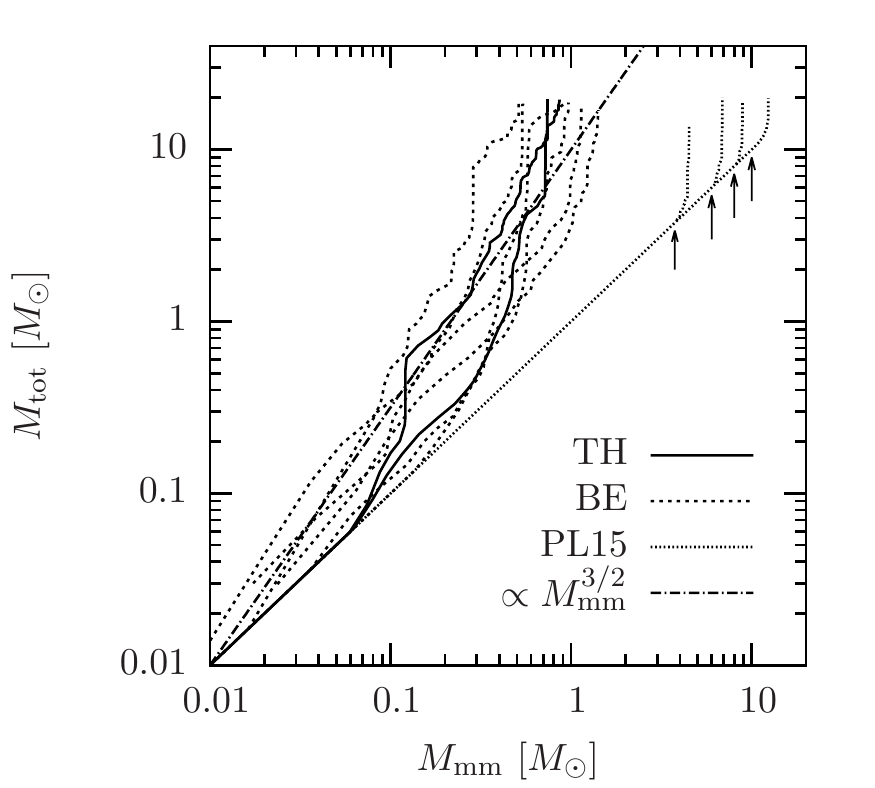}
  \caption{Relation of the most massive star in the cluster to the total cluster mass. Note that in the PL15 case, secondary sink particles form very late in the simulation (indicated by the arrows), which explains the large range in mass with a slope of unity.}
  \label{fig:mm-mass-relations}
\end{figure}

Depending on the formation time, the formation location and the accretion rate over time, the mass of the most massive protostar in relation to the mass of the other objects or the entire ensemble of protostars in the cluster might differ for different setups and initial conditions. Figure~\ref{fig:mm-mass-relations} shows the mass of the most massive sink particle, $M_\mathrm{mm}$, in relation to the total mass confined in sink particles, $M_\mathrm{tot}$. The setups with strong initial mass concentrations (PL15) form an early protostar that stays the only sink particle for a significant part of the total simulation. After the formation of subsequent protostars, the mass of the central massive one stays almost constant due to the starvation effect. The mass relations of all PL15 setups therefore bend upwards. In the setups with initially flat density distribution, the mass of the most massive protostar roughly follows a relation $M_\mathrm{mm} \propto M_\mathrm{tot}^{2/3}$, but with remarkable scatter. The fact that the difference in mass between the most massive and the other stars is much less, leads to a stronger impact of $N$-body interactions on the location of the most massive protostar within the cluster. While the very massive central stars in the PL15 runs remain within the accretion shielded area of the cluster, the most massive stars in the BE runs and in the subclusters of the TH setups follow larger orbits where they leave and reenter the shielded area alternately. Whenever the most massive sink particle gains further mass, either by leaving the accretion shielded area or experiencing episodic accretion, the curve flattens. Entering the accretion shielded volume of the cluster, the most massive object gets starved of material and its accretion stops. As the whole cluster continues to grow, the curves bend up.

\section{Discussion}

Two accretion models are widely discussed in the literature: 1) the monolithic collapse and 2) competitive accretion and fragmentation induced starvation, where mutual dynamical interactions of accreting protostars play a key role in shaping the final mass distribution. The monolithic collapse model is based on the apparent similarity of the core mass function \citep{Motte98,Johnstone00,Johnstone06,Lada08} and the stellar initial mass function \citep{Kroupa02Sci,Chabrier03}. It relies on a direct mapping of cores to stars with a constant efficiency factor for the collapse of a single core into a single stars or at most a binary system. A major caveat in this model is the collapse of very massive cores. The amount of turbulent energy deposited in these cores is likely to cause the core to fragment into many objects. This work and especially Paper~I demonstrate that only strongly concentrated density profiles may prevent further fragmentation in an isothermal environment. Radiation feedback tends to reduce the degree of fragmentation, but still can not prevent the collapsing core from fragmenting into many protostars \citep{Krumholz07,Bate09,Peters10a}. Likewise magnetic fields support the agglomeration of larger masses. However, even magnetised cores fragment into smaller objects \citep{Ziegler05,HennebelleTeyssier2008,Banerjee09,Buerzle11,Peters11,Hennebelle11,SeifriedEtAl2011}. The degree of fragmentation and the resulting dynamical accretion in our simulated dense clusters significantly differ from the monolithic collapse model. The fragmentation is suppressed only for highly concentrated density profiles and weak initial turbulent motions (see Paper~I). Apart from the fragmentation of the gas, the assumption of a monolithic collapse leads to a time-scale problem that is likely to destroy the apparent similarity of the IMF and the core mass function \citep{Clark07,Smith08}.

In competitive accretion \citep{Bonnell01, Bonnell01b, Bonnell02, Bate05}, the cloud first fragments down to objects with masses close to the opacity limit, which then start to accrete and build up the initial mass function. The accretion rates are mainly determined by the position of the protostar in the cluster. The gravitational potential funnels gas to the centre of the cluster, leading to higher accretion rates onto the most central objects, located close to the centre of the potential well. Thus, central protostars can grow to the most massive ones. The more massive a star, the faster it can dynamically migrate to the centre of mass, continuously ensuring a gravitationally privileged position. Assuming a negligible impact of the surrounding protostars on the global mass accretion, the most massive stars continue to accrete at a higher accretion rate. As soon as the impact of further fragmentation influences the global accretion process, the central accretion rates may vary significantly, as it is described in the fragmentation induced starvation model. In competitive accretion, the mass of the most massive star is related to the total stellar mass as $M_\mathrm{mm} \propto M_\mathrm{tot}^{2/3}$ \citep{Bonnell04}. This relation was assumed to be an indicator for this accretion model \citep{Krumholz09sfa} and is in agreement with observations \citep{Weidner06, Weidner10}. However, as we discuss below, fragmentation induced starvation can lead to the same behaviour. This relation is therefore not a unique signpost of competitive accretion.

Intermediate scenarios between the two extreme models are also reported \citep[e.g.][]{Peretto06,Wang10}, in which the formation model depends on the mass of the star. Low-mass objects form via local collapse, while massive protostars and their dense proximity accumulate gas from the global environment.

Recent numerical models of star cluster formation indicate that the competitive accretion model in its original flavour \citep{KlessenBurkert2000, KlessenBurkert2001, Bonnell01, Bonnell01b, Bonnell02, Bate05} needs reinterpretation. The suggestion of higher central accretion rates due to the gravitationally privileged position changes once the formed fragments around the centre of the collapsing area have a significant impact on the global gas accretion process. The infalling gas may be branched off and either be accreted onto these objects or dynamically redirected before reaching the centre of the cluster. \citet{Peters10c} found a strong impact of subsequent fragmentation on the accretion flow in a $1000\,M_\odot$ rotating core including radiative feedback, but without initial turbulence. The accretion rate onto the central object drops significantly. Instead of growing further, the central protostar is starved of material. Despite this significantly different accretion behaviour, they also found masses in the simulation in good agreement with the relation $M_\mathrm{mm} \propto M_\mathrm{tot}^{2/3}$, ruling out this relation as evidence for competitive accretion. As they do not apply turbulent motions, but only solid body rotation to the initial cloud, they focus on the starvation effect in disc-like conditions rather than in filamentary structures. Due to the missing turbulence, their disc-like structure does not evolve into a volume-filling cluster, but remains a flattened structure. A disc provides a smaller effective area through which gas can flow towards the centre. In a cluster with random protostellar trajectories, the effective accretion area is much larger. Therefore, from simple geometrical arguments, it is much more difficult to shield the central area from accretion flows.

However, we find that protostars in the outer region of the cluster form along the densest part of extended filaments, which allows them to accrete much more mass than in an idealised spherically symmetric setup. In addition to that, we observe significantly smaller relative velocities between the protostars and the immediate surrounding gas for protostars at larger distances from the centre of the cluster. This increases their geometrical accretion cross section, expressed by the Bondi-Hoyle accretion radius. Altogether, the accretion in the central region of the cluster is influenced by efficiently shielding protostars located at the position of the accretion flow. As a result, even a relatively small number of surrounding stars is sufficient to provide an efficient starvation effect in a volume-filling cluster environment.

One caveat of the presented simulations is the missing feedback from the stars that are formed first in the clusters. The gas stays isothermal and therefore tends to fragment much more quickly than with heating of the gas. Including this effect, the number of protostars is expected to be smaller. The starvation effect on the central objects in the protostellar cluster might thus be overestimated, especially in a turbulent volume-filling cluster, where gas can fall into the central region from all directions. However, \citet{Peters10c} found that the number of protostars in simulations including radiative feedback is roughly half of the number of protostars in an isothermal calculation. As the total number of sink particles in our simulations is much larger than $100$, and additionally, the starvation effect can already be seen with a relatively small number of competing protostars, we do not expect the shielding effect to completely vanish in the non-isothermal case. As long as there is some fragmentation, the fragments are likely to shield the central massive star from accretion, irrespective of whether radiation feedback is included or not. Likewise do magnetic fields tend to reduce the degree of fragmentation, but still do not prevent the cloud from fragmenting \citep{Hennebelle11,Peters11}.

Finally, we want to point to recent studies by \citet{KruijssenEtAl2011}. They analysed the substructure within clusters as well as the dynamical state of the stellar cluster when gas expulsion becomes important, i.e., at a slightly later stage of the evolution of the cluster. Analysing the simulations of \citet{Bonnell03,BonnellEtAl2008}, they find that the stellar system quickly reaches a globally virialised state if the gas potential is excluded and the stellar system is followed with pure $N$-body dynamics. Their results support the evolutionary picture of the formation of protostars that we see in our simulations. New protostars that form at larger radii from the centre of the cluster in gas-dominated regions have sub-virial velocities. As soon as they decouple from the gas motion and move to the central gas-poor environment, they quickly virialise.

\section{Summary and Conclusion}
We performed simulations of collapsing molecular gas clouds with a total mass of $100\,M_\odot$ and a diameter of $0.2\,\mathrm{pc}$. We varied the initial density profile as well as the turbulent supersonic velocity field and analysed the fragmentation process and the accretion onto nascent protostars. Most clouds, especially the ones with initially less centrally concentrated density profiles undergo fragmentation and form compact stellar clusters. We studied the accretion process in these clusters with the following conclusions:
\begin{itemize}
\item All clusters show strong and fast dynamical interactions between the protostars. During the formation time of the cluster, the protostars complete several orbits in the parental cluster. The mass accretion process can thus not be described by a monolithic collapse model.
\item Fragmentation and formation of multiple protostars strongly influence the subsequent accretion flow in the entire cluster. Gas that falls towards the centre of the cluster is efficiently accreted onto protostars that are located at larger radii from the cluster centre. As a consequence, the central region is effectively shielded from further accretion and none of the central objects can sustain its initially high accretion rate. In all cases the observed decrease of the central accretion rate is due to the efficient shielding by secondary protostars. A significant starvation effect due to angular momentum and a resulting radial barrier can be excluded.
\item In setups with initial flat density distributions, the mass of the most massive protostar, $M_\mathrm{mm}$, scales with the total cluster mass, $M_\mathrm{tot}$, like $M_\mathrm{mm} \propto M_\mathrm{tot}^{2/3}$, as originally proposed as a probe for competitive accretion. This relation is a common feature to both dynamical accretion models and cannot be used to distinguish between competitive accretion and fragmentation induced starvation, a conclusion that was already reached by \citet{Peters10b}.
\item The accretion process does not differ between density profiles that form only one main cluster (power-law profiles, Bonnor-Ebert density distribution) and setups that form multiple subclusters (uniform density profile). In both, the central clusters as well as the disconnected subclusters, we find fragmentation induced starvation to work in the central region of the cluster.
\end{itemize}

\section*{Acknowledgement}
We thank Thomas Peters for many interesting discussions and comments as well as the referee, Ian Bonnell, for very useful comments and suggestions to improve the depiction of the relevant physics. P.G.~acknowledges supercomputer grants at the J\"{u}lich supercomputing centre (NIC~3433) and at the CASPUR centre (cmp09-849). P.G.~and C.F.~are grateful for financial support from the International Max Planck Research School for Astronomy and Cosmic Physics (IMPRS-A) and the Heidelberg Graduate School of Fundamental Physics (HGSFP), funded by the Excellence Initiative of the Deutsche Forschungsgemeinschaft (DFG) under grant GSC~129/1. C.F., R.B.~and R.S.K.~acknowledge financial support from the Landesstiftung Baden-W\"{u}rttemberg via their program \emph{Internationale Spitzenforschung~II} (grant P-LS-SPII/18) and from the German Bundesministerium f\"{u}r Bildung und Forschung via the ASTRONET project STAR FORMAT (grant 05A09VHA). C.F.~furthermore acknowledges funding from the European Research Council under the European Community’s Seventh Framework Programme (FP7/2007-2013 Grant Agreement no. 247060) and from the Australian Research Council for a Discovery Projects Fellowship (grant no.~DP110102191). P.G. and R.B.~acknowledge funding of Emmy Noether grant BA~3706/1-1 by the DFG. R.B.~is furthermore thankful for subsidies from the FRONTIER initiative of the University of Heidelberg. R.S.K.~acknowledges subsidies from the DFG under grants no.~KL1358/10, and KL1358/11, the Sonderforschungsbereich SFB~881 \emph{The Milky Way System} as well as from a Frontier grant of Heidelberg University sponsored by the German Excellence Initiative. This work was supported in part by the U.S.~Department of Energy contract no.~DEAC-02-76SF00515. The FLASH code was developed in part by the DOE-supported Alliances Center for Astrophysical Thermonuclear Flashes (ASC) at the University of Chicago.

\clearpage
\newpage
% \bibliographystyle{mn2e}
% Use apj.bst instead of mn2e.bst, because only apj.bst distinguishes between 2010a and 2010b references automatically.
\bibliographystyle{apj}
\bibliography{astro.bib}

\begin{thebibliography}{}

\bibitem[\protect\citeauthoryear{Banerjee \& Pudritz}{Banerjee \&
  Pudritz}{2006}]{Banerjee06}
Banerjee, R.,  \& Pudritz, R.~E. 2006, {\apj}, 641, 949

\bibitem[\protect\citeauthoryear{{Banerjee} et~al.}{{Banerjee}
  et~al.}{2009}]{Banerjee09}
{Banerjee}, R., {V{\'a}zquez-Semadeni}, E., {Hennebelle}, P.,  \& {Klessen},
  R.~S. 2009, \mnras, 398, 1082

\bibitem[\protect\citeauthoryear{{Bastian}, {Covey}, \& {Meyer}}{{Bastian}
  et~al.}{2010}]{BastianEtAl2010}
{Bastian}, N., {Covey}, K.~R.,  \& {Meyer}, M.~R. 2010, \araa, 48, 339

\bibitem[\protect\citeauthoryear{{Bate}}{{Bate}}{2009}]{Bate09}
{Bate}, M.~R. 2009, \mnras, 392, 590

\bibitem[\protect\citeauthoryear{{Bate} \& {Bonnell}}{{Bate} \&
  {Bonnell}}{2005}]{Bate05}
{Bate}, M.~R.,  \& {Bonnell}, I.~A. 2005, \mnras, 356, 1201

\bibitem[\protect\citeauthoryear{{Baumgardt} \& {Klessen}}{{Baumgardt} \&
  {Klessen}}{2011}]{Baumgardt11}
{Baumgardt}, H.,  \& {Klessen}, R.~S. 2011, \mnras, 413, 1810

\bibitem[\protect\citeauthoryear{{Bertoldi}}{{Bertoldi}}{1989}]{Bertoldi89}
{Bertoldi}, F. 1989, \apj, 346, 735

\bibitem[\protect\citeauthoryear{{Bertoldi} \& {McKee}}{{Bertoldi} \&
  {McKee}}{1990}]{Bertoldi90}
{Bertoldi}, F.,  \& {McKee}, C.~F. 1990, \apj, 354, 529

\bibitem[\protect\citeauthoryear{{Bondi}}{{Bondi}}{1952}]{Bondi52}
{Bondi}, H. 1952, \mnras, 112, 195

\bibitem[\protect\citeauthoryear{{Bondi} \& {Hoyle}}{{Bondi} \&
  {Hoyle}}{1944}]{Bondi44}
{Bondi}, H.,  \& {Hoyle}, F. 1944, \mnras, 104, 273

\bibitem[\protect\citeauthoryear{{Bonnell} \& {Bate}}{{Bonnell} \&
  {Bate}}{2002}]{Bonnell02}
{Bonnell}, I.~A.,  \& {Bate}, M.~R. 2002, \mnras, 336, 659

\bibitem[\protect\citeauthoryear{{Bonnell} et~al.}{{Bonnell}
  et~al.}{2001a}]{Bonnell01}
{Bonnell}, I.~A., {Bate}, M.~R., {Clarke}, C.~J.,  \& {Pringle}, J.~E. 2001a,
  \mnras, 323, 785

\bibitem[\protect\citeauthoryear{{Bonnell}, {Bate}, \& {Vine}}{{Bonnell}
  et~al.}{2003}]{Bonnell03}
{Bonnell}, I.~A., {Bate}, M.~R.,  \& {Vine}, S.~G. 2003, \mnras, 343, 413

\bibitem[\protect\citeauthoryear{{Bonnell}, {Clark}, \& {Bate}}{{Bonnell}
  et~al.}{2008}]{BonnellEtAl2008}
{Bonnell}, I.~A., {Clark}, P.,  \& {Bate}, M.~R. 2008, \mnras, 389, 1556

\bibitem[\protect\citeauthoryear{{Bonnell} et~al.}{{Bonnell}
  et~al.}{2001b}]{Bonnell01b}
{Bonnell}, I.~A., {Clarke}, C.~J., {Bate}, M.~R.,  \& {Pringle}, J.~E. 2001b,
  \mnras, 324, 573

\bibitem[\protect\citeauthoryear{{Bonnell}, {Vine}, \& {Bate}}{{Bonnell}
  et~al.}{2004}]{Bonnell04}
{Bonnell}, I.~A., {Vine}, S.~G.,  \& {Bate}, M.~R. 2004, \mnras, 349, 735

\bibitem[\protect\citeauthoryear{{Bressert} et~al.}{{Bressert}
  et~al.}{2010}]{BressertEtAl2010}
{Bressert}, E., et~al. 2010, \mnras, 409, L54

\bibitem[\protect\citeauthoryear{{B{\"u}rzle} et~al.}{{B{\"u}rzle}
  et~al.}{2011}]{Buerzle11}
{B{\"u}rzle}, F., {Clark}, P.~C., {Stasyszyn}, F., {Greif}, T., {Dolag}, K.,
  {Klessen}, R.~S.,  \& {Nielaba}, P. 2011, \mnras, 412, 171

\bibitem[\protect\citeauthoryear{{Chabrier}}{{Chabrier}}{2003}]{Chabrier03}
{Chabrier}, G. 2003, \pasp, 115, 763

\bibitem[\protect\citeauthoryear{{Clark}, {Klessen}, \& {Bonnell}}{{Clark}
  et~al.}{2007}]{Clark07}
{Clark}, P.~C., {Klessen}, R.~S.,  \& {Bonnell}, I.~A. 2007, \mnras, 379, 57

\bibitem[\protect\citeauthoryear{{Colella} \& {Woodward}}{{Colella} \&
  {Woodward}}{1984}]{Colella84}
{Colella}, P.,  \& {Woodward}, P.~R. 1984, Journal of Computational Physics,
  54, 174

\bibitem[\protect\citeauthoryear{{Commer{\c c}on} et~al.}{{Commer{\c c}on}
  et~al.}{2010}]{Commercon10}
{Commer{\c c}on}, B., {Hennebelle}, P., {Audit}, E., {Chabrier}, G.,  \&
  {Teyssier}, R. 2010, \aap, 510, L3

\bibitem[\protect\citeauthoryear{{Elmegreen} \& {Lada}}{{Elmegreen} \&
  {Lada}}{1977}]{Elmegreen77}
{Elmegreen}, B.~G.,  \& {Lada}, C.~J. 1977, \apj, 214, 725

\bibitem[\protect\citeauthoryear{{Federrath} et~al.}{{Federrath}
  et~al.}{2010a}]{Federrath10a}
{Federrath}, C., {Banerjee}, R., {Clark}, P.~C.,  \& {Klessen}, R.~S. 2010a,
  \apj, 713, 269

\bibitem[\protect\citeauthoryear{{Federrath}, {Klessen}, \&
  {Schmidt}}{{Federrath} et~al.}{2008}]{Federrath08}
{Federrath}, C., {Klessen}, R.~S.,  \& {Schmidt}, W. 2008, \apjl, 688, L79

\bibitem[\protect\citeauthoryear{{Federrath} et~al.}{{Federrath}
  et~al.}{2010b}]{Federrath10b}
{Federrath}, C., {Roman-Duval}, J., {Klessen}, R.~S., {Schmidt}, W.,  \& {Mac
  Low}, M. 2010b, \aap, 512, A81

\bibitem[\protect\citeauthoryear{{Fryxell} et~al.}{{Fryxell}
  et~al.}{2000}]{FLASH00}
{Fryxell}, B., et~al. 2000, \apjs, 131, 273

\bibitem[\protect\citeauthoryear{{Girichidis} et~al.}{{Girichidis}
  et~al.}{2011}]{Girichidis11a}
{Girichidis}, P., {Federrath}, C., {Banerjee}, R.,  \& {Klessen}, R.~S. 2011,
  \mnras, 413, 2741

\bibitem[\protect\citeauthoryear{{Hennebelle} \& {Ciardi}}{{Hennebelle} \&
  {Ciardi}}{2009}]{HennebelleCiardi09}
{Hennebelle}, P.,  \& {Ciardi}, A. 2009, \aap, 506, L29

\bibitem[\protect\citeauthoryear{{Hennebelle} et~al.}{{Hennebelle}
  et~al.}{2011}]{Hennebelle11}
{Hennebelle}, P., {Commer{\c c}on}, B., {Joos}, M., {Klessen}, R.~S.,
  {Krumholz}, M., {Tan}, J.~C.,  \& {Teyssier}, R. 2011, \aap, 528, A72

\bibitem[\protect\citeauthoryear{{Hennebelle} \& {Fromang}}{{Hennebelle} \&
  {Fromang}}{2008}]{Hennebelle08}
{Hennebelle}, P.,  \& {Fromang}, S. 2008, \aap, 477, 9

\bibitem[\protect\citeauthoryear{{Hennebelle} \& {Teyssier}}{{Hennebelle} \&
  {Teyssier}}{2008}]{HennebelleTeyssier2008}
{Hennebelle}, P.,  \& {Teyssier}, R. 2008, \aap, 477, 25

\bibitem[\protect\citeauthoryear{{Heyer} \& {Brunt}}{{Heyer} \&
  {Brunt}}{2004}]{Heyer04}
{Heyer}, M.~H.,  \& {Brunt}, C.~M. 2004, \apjl, 615, L45

\bibitem[\protect\citeauthoryear{{Johnstone}, {Matthews}, \&
  {Mitchell}}{{Johnstone} et~al.}{2006}]{Johnstone06}
{Johnstone}, D., {Matthews}, H.,  \& {Mitchell}, G.~F. 2006, \apj, 639, 259

\bibitem[\protect\citeauthoryear{{Johnstone} et~al.}{{Johnstone}
  et~al.}{2000}]{Johnstone00}
{Johnstone}, D., {Wilson}, C.~D., {Moriarty-Schieven}, G., {Joncas}, G.,
  {Smith}, G., {Gregersen}, E.,  \& {Fich}, M. 2000, \apj, 545, 327

\bibitem[\protect\citeauthoryear{{Klessen}}{{Klessen}}{2001}]{Klessen01}
{Klessen}, R.~S. 2001, \apj, 556, 837

\bibitem[\protect\citeauthoryear{{Klessen} \& {Burkert}}{{Klessen} \&
  {Burkert}}{2000}]{KlessenBurkert2000}
{Klessen}, R.~S.,  \& {Burkert}, A. 2000, \apjs, 128, 287

\bibitem[\protect\citeauthoryear{{Klessen} \& {Burkert}}{{Klessen} \&
  {Burkert}}{2001}]{KlessenBurkert2001}
{Klessen}, R.~S.,  \& {Burkert}, A. 2001, \apj, 549, 386

\bibitem[\protect\citeauthoryear{{Klessen}, {Heitsch}, \& {Mac Low}}{{Klessen}
  et~al.}{2000}]{KlessenHeitsch00}
{Klessen}, R.~S., {Heitsch}, F.,  \& {Mac Low}, M. 2000, \apj, 535, 887

\bibitem[\protect\citeauthoryear{{Kroupa}}{{Kroupa}}{2001}]{Kroupa01}
{Kroupa}, P. 2001, \mnras, 322, 231

\bibitem[\protect\citeauthoryear{{Kroupa}}{{Kroupa}}{2002}]{Kroupa02Sci}
{Kroupa}, P. 2002, Science, 295, 82

\bibitem[\protect\citeauthoryear{{Kruijssen} et~al.}{{Kruijssen}
  et~al.}{2011}]{KruijssenEtAl2011}
{Kruijssen}, J.~M.~D., {Maschberger}, T., {Moeckel}, N., {Clarke}, C.~J.,
  {Bastian}, N.,  \& {Bonnell}, I.~A. 2011, \mnras, 1779

\bibitem[\protect\citeauthoryear{{Krumholz} \& {Bonnell}}{{Krumholz} \&
  {Bonnell}}{2009}]{Krumholz09sfa}
{Krumholz}, M.~R.,  \& {Bonnell}, I.~A. 2009, {Models for the formation of
  massive stars}, ed. {Chabrier, G.} (Cambridge University Press), 288

\bibitem[\protect\citeauthoryear{{Krumholz}, {Klein}, \& {McKee}}{{Krumholz}
  et~al.}{2007}]{Krumholz07}
{Krumholz}, M.~R., {Klein}, R.~I.,  \& {McKee}, C.~F. 2007, \apj, 656, 959

\bibitem[\protect\citeauthoryear{{Lada} et~al.}{{Lada} et~al.}{2003}]{Lada03}
{Lada}, C.~J., {Bergin}, E.~A., {Alves}, J.~F.,  \& {Huard}, T.~L. 2003, \apj,
  586, 286

\bibitem[\protect\citeauthoryear{{Lada} et~al.}{{Lada} et~al.}{2008}]{Lada08}
{Lada}, C.~J., {Muench}, A.~A., {Rathborne}, J., {Alves}, J.~F.,  \&
  {Lombardi}, M. 2008, \apj, 672, 410

\bibitem[\protect\citeauthoryear{{Larson}}{{Larson}}{1981}]{Larson81}
{Larson}, R.~B. 1981, \mnras, 194, 809

\bibitem[\protect\citeauthoryear{{Mac Low} \& {Klessen}}{{Mac Low} \&
  {Klessen}}{2004}]{MacLow04}
{Mac Low}, M.-M.,  \& {Klessen}, R.~S. 2004, Reviews of Modern Physics, 76, 125

\bibitem[\protect\citeauthoryear{{McKee} \& {Ostriker}}{{McKee} \&
  {Ostriker}}{2007}]{McKee07}
{McKee}, C.~F.,  \& {Ostriker}, E.~C. 2007, \araa, 45, 565

\bibitem[\protect\citeauthoryear{{McKee} \& {Tan}}{{McKee} \&
  {Tan}}{2002}]{McKee02}
{McKee}, C.~F.,  \& {Tan}, J.~C. 2002, \nat, 416, 59

\bibitem[\protect\citeauthoryear{{McKee} \& {Tan}}{{McKee} \&
  {Tan}}{2003}]{McKee03}
{McKee}, C.~F.,  \& {Tan}, J.~C. 2003, \apj, 585, 850

\bibitem[\protect\citeauthoryear{{Motte}, {Andre}, \& {Neri}}{{Motte}
  et~al.}{1998}]{Motte98}
{Motte}, F., {Andre}, P.,  \& {Neri}, R. 1998, \aap, 336, 150

\bibitem[\protect\citeauthoryear{{Olson} et~al.}{{Olson}
  et~al.}{1999}]{PARAMESH99}
{Olson}, K.~M., {MacNeice}, P., {Fryxell}, B., {Ricker}, P., {Timmes}, F.~X.,
  \& {Zingale}, M. 1999, Bulletin of the American Astronomical Society, 31,
  1430

\bibitem[\protect\citeauthoryear{{Peretto}, {Andr{\'e}}, \&
  {Belloche}}{{Peretto} et~al.}{2006}]{Peretto06}
{Peretto}, N., {Andr{\'e}}, P.,  \& {Belloche}, A. 2006, \aap, 445, 979

\bibitem[\protect\citeauthoryear{{Peters} et~al.}{{Peters}
  et~al.}{2011}]{Peters11}
{Peters}, T., {Banerjee}, R., {Klessen}, R.~S.,  \& {Mac Low}, M. 2011, \apj,
  729, 72

\bibitem[\protect\citeauthoryear{{Peters} et~al.}{{Peters}
  et~al.}{2010a}]{Peters10a}
{Peters}, T., {Banerjee}, R., {Klessen}, R.~S., {Mac Low}, M.,
  {Galv{\'a}n-Madrid}, R.,  \& {Keto}, E.~R. 2010a, \apj, 711, 1017

\bibitem[\protect\citeauthoryear{{Peters} et~al.}{{Peters}
  et~al.}{2010b}]{Peters10c}
{Peters}, T., {Klessen}, R.~S., {Mac Low}, M.,  \& {Banerjee}, R. 2010b, \apj,
  725, 134

\bibitem[\protect\citeauthoryear{{Peters} et~al.}{{Peters}
  et~al.}{2010c}]{Peters10b}
{Peters}, T., {Mac Low}, M., {Banerjee}, R., {Klessen}, R.~S.,  \& {Dullemond},
  C.~P. 2010c, \apj, 719, 831

\bibitem[\protect\citeauthoryear{{Price} \& {Bate}}{{Price} \&
  {Bate}}{2007}]{Price07}
{Price}, D.~J.,  \& {Bate}, M.~R. 2007, \apss, 311, 75

\bibitem[\protect\citeauthoryear{{Scalo}}{{Scalo}}{1998}]{Scalo98}
{Scalo}, J. 1998, in Astronomical Society of the Pacific Conference Series,
  Vol. 142, The Stellar Initial Mass Function (38th Herstmonceux Conference),
  ed. {G.~Gilmore \& D.~Howell}, 201

\bibitem[\protect\citeauthoryear{{Scalo}}{{Scalo}}{1986}]{Scalo86}
{Scalo}, J.~M. 1986, Fundamentals of Cosmic Physics, 11, 1

\bibitem[\protect\citeauthoryear{{Schmeja} \& {Klessen}}{{Schmeja} \&
  {Klessen}}{2004}]{SchmejaKlessen04}
{Schmeja}, S.,  \& {Klessen}, R.~S. 2004, \aap, 419, 405

\bibitem[\protect\citeauthoryear{{Seifried} et~al.}{{Seifried}
  et~al.}{2011}]{SeifriedEtAl2011}
{Seifried}, D., {Banerjee}, R., {Klessen}, R.~S., {Duffin}, D.,  \& {Pudritz},
  R.~E. 2011, \mnras, 417, 1054

\bibitem[\protect\citeauthoryear{{Smith}, {Clark}, \& {Bonnell}}{{Smith}
  et~al.}{2008}]{Smith08}
{Smith}, R.~J., {Clark}, P.~C.,  \& {Bonnell}, I.~A. 2008, \mnras, 391, 1091

\bibitem[\protect\citeauthoryear{{Smith}, {Longmore}, \& {Bonnell}}{{Smith}
  et~al.}{2009}]{Smith09}
{Smith}, R.~J., {Longmore}, S.,  \& {Bonnell}, I. 2009, \mnras, 400, 1775

\bibitem[\protect\citeauthoryear{{Testi} \& {Sargent}}{{Testi} \&
  {Sargent}}{1998}]{Testi98}
{Testi}, L.,  \& {Sargent}, A.~I. 1998, \apjl, 508, L91

\bibitem[\protect\citeauthoryear{{Truelove} et~al.}{{Truelove}
  et~al.}{1997}]{Truelove97}
{Truelove}, J.~K., {Klein}, R.~I., {McKee}, C.~F., {Holliman}, J.~H., {Howell},
  L.~H.,  \& {Greenough}, J.~A. 1997, \apjl, 489, L179

\bibitem[\protect\citeauthoryear{{Wang} et~al.}{{Wang} et~al.}{2010}]{Wang10}
{Wang}, P., {Li}, Z., {Abel}, T.,  \& {Nakamura}, F. 2010, \apj, 709, 27

\bibitem[\protect\citeauthoryear{{Weidner} \& {Kroupa}}{{Weidner} \&
  {Kroupa}}{2006}]{Weidner06}
{Weidner}, C.,  \& {Kroupa}, P. 2006, \mnras, 365, 1333

\bibitem[\protect\citeauthoryear{{Weidner}, {Kroupa}, \& {Bonnell}}{{Weidner}
  et~al.}{2010}]{Weidner10}
{Weidner}, C., {Kroupa}, P.,  \& {Bonnell}, I.~A.~D. 2010, \mnras, 401, 275

\bibitem[\protect\citeauthoryear{{Whitworth} et~al.}{{Whitworth}
  et~al.}{1994}]{Whitworth94}
{Whitworth}, A.~P., {Bhattal}, A.~S., {Chapman}, S.~J., {Disney}, M.~J.,  \&
  {Turner}, J.~A. 1994, \mnras, 268, 291

\bibitem[\protect\citeauthoryear{{Ziegler}}{{Ziegler}}{2005}]{Ziegler05}
{Ziegler}, U. 2005, \aap, 435, 385

\bibitem[\protect\citeauthoryear{{Zinnecker} \& {Yorke}}{{Zinnecker} \&
  {Yorke}}{2007}]{Zinnecker07}
{Zinnecker}, H.,  \& {Yorke}, H.~W. 2007, \araa, 45, 481

\end{thebibliography}
\clearpage
\begin{appendix}
  \section{Angular Momentum}
  \begin{figure}
    \includegraphics[width=8cm]{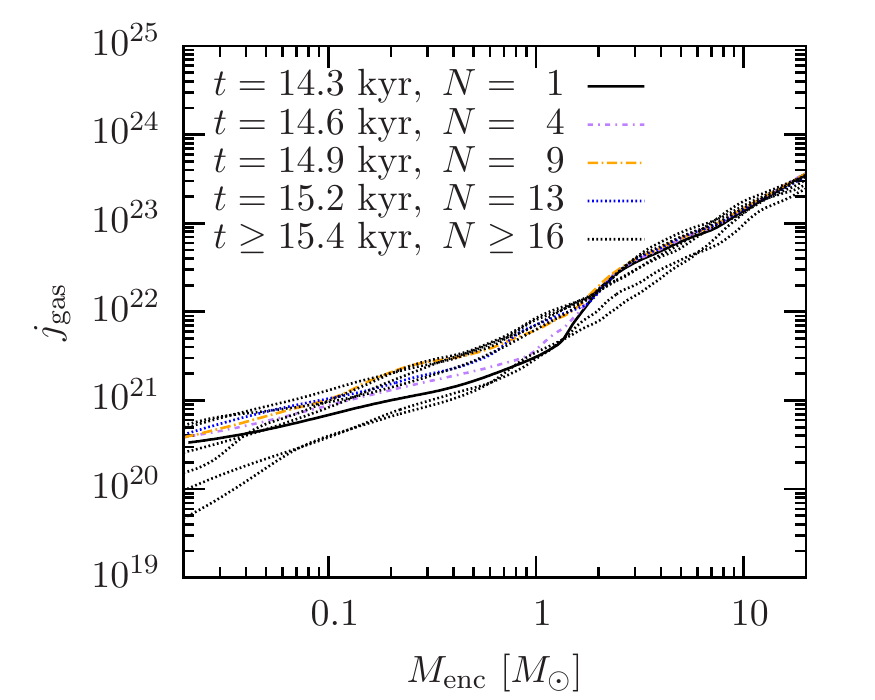}
    \caption{Specific angular momentum of the gas as a function of enclosed mass for PL15-m-2. The plot shows temporal changes of the angular momentum of the gas during the formation of the cluster.}
    \label{fig:PL15m2-angmom-Menc}
  \end{figure}
  \begin{figure}
    \includegraphics[width=8cm]{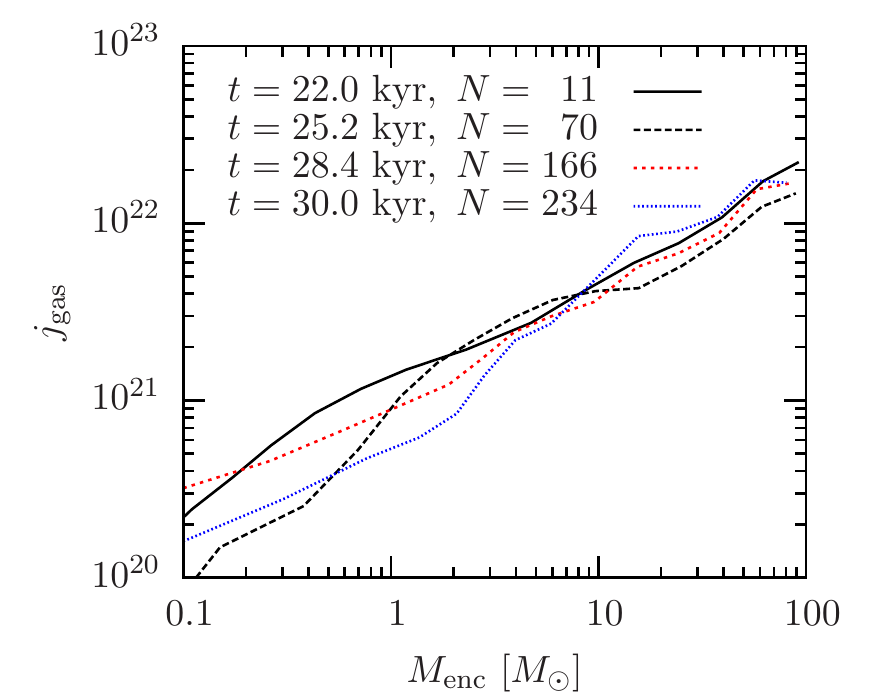}
    \caption{Same as figure~\ref{fig:PL15m2-angmom-Menc} but for setup BE-m-2.}
    \label{fig:BEm2-angmom-Menc}
  \end{figure}
  In order to see the time evolution of the angular momentum, we show the specific angular momentum as a function of enclosed mass for the setups PL15-m-2 and BE-m-2 (figures~\ref{fig:PL15m2-angmom-Menc} and \ref{fig:BEm2-angmom-Menc}). In both cases the angular momentum distribution is exposed to variations of the order of a few up to almost an order of magnitude due to the turbulent interactions and the gas accretion onto protostars. Due to the unstructured motions of the protostars and the random character of the turbulence, the variations do not show systematic changes over time, i.e., the gas in the cluster can quickly gain angular momentum due to accretion streams or lose it via shocks and dissipation.

\end{appendix}

\end{document}